%% file: main.tex
\documentclass[seceq,twocolumn]{jpsj2}
\usepackage{amsmath, amsfonts, amsthm, amssymb}
\usepackage{graphicx}
\usepackage{bm}
\usepackage{accents}
\usepackage{enumerate}
\usepackage{color}
\usepackage{arydshln}
\usepackage{array}
\usepackage{colortbl}
\usepackage{ulem}
\usepackage{url}
\definecolor{lightgray}{rgb}{0.75,0.75,0.75}
\definecolor{darkgray}{rgb}{0.25,0.25,0.25}

\newcommand{\bhline}[1]{\noalign{\hrule height #1}}
\newcommand{\bvline}[1]{\vrule width #1}

\def\graphwidthsingle{0.325\textwidth}

\def\Eqs{eqs.}
\def\Eq{eq.}

\newcommand{\Ns}{\ensuremath{N_\mathrm{s}}}
\newcommand{\Ne}{\ensuremath{N_\mathrm{e}}}

\definecolor{red}{rgb}{0.5, 0.0, 0.0}

\def\coloronline{(Color online)~}
\def\journal#1#2#3#4#5{#1: #2 {\bf #3} (#5) #4.}

\def\arXiv#1#2{#1: cond-mat/#2.}
\def\book#1#2#3{#1: \textit{#2}, #3.}

\def\PR{Phys.\ Rev.}
\def\PRL{Phys.\ Rev.\ Lett.}
\def\PRB{Phys.\ Rev.\ B}

\def\RMP{Rev.\ Mod.\ Phys.}

\def\JPC{J. Phys. C}
\def\EL{Europhys. Lett.}
\def\JPCM{J. Phys.: Condens. Matter}

\renewcommand{\[}{\begin{equation}}
\renewcommand{\]}{\end{equation}}


\def\runauthor{\textsc{H.~Shinaoka} \textit{et al.}}
\makeatletter
\def\@typeset{}
\makeatother

\begin{document}
\title{Single-particle excitations under coexisting electron correlation and disorder: a numerical study of the Anderson-Hubbard model}

\author{Hiroshi {\sc SHINAOKA}$^{1}$\thanks{E-mail: shinaoka@solis.t.u-tokyo.ac.jp} and Masatoshi {\sc IMADA}$^{1,2}$}
\inst{$^{1}$ Department of Applied Physics, The University of Tokyo, Tokyo 113-8656\\
$^{2}$ CREST, JST, 7-3-1 Hongo, Bunkyo-ku, Tokyo 113-8656} 
\recdate{\today}

\date{\today}
\abst{
  \input{abst01.tex}
}

\kword{electron correlation, disorder, Anderson-Hubbard model, single-particle density of states, soft gap, variable-range hopping}

	\maketitle
  \input{chap01.tex}
  \input{method.tex}
  \input{numerical.tex}
  \input{theory.tex}
  \input{transport.tex}
  \input{conclusion.tex}
  \input{ackn01.tex}
	\appendix
  \input{ap01.tex}

\input{bib01.tex}
\end{document}

%% file: abst01.tex
Interplay of electron correlation and randomness is studied by using the Anderson-Hubbard model within the Hartree-Fock approximation. Under the coexistence of short-range interaction and diagonal disorder, we obtain the ground-state phase diagram in three dimensions, which includes an antiferromagnetic insulator, an antiferromagnetic metal, a paramagnetic insulator (Anderson-localized insulator) and a paramagnetic metal. Although only the short-range interaction is present in this model, we find unconventional soft gaps in the insulating phases irrespective of electron filling, spatial dimensions and long-range order, where the single-particle density of states (DOS) vanishes with a power-law scaling in one dimension (1D) or even faster in two dimensions (2D) and three dimensions (3D) toward the Fermi energy. We call it \textit{soft Hubbard gap}. Moreover, exact-diagonalization results in 1D support the formation of the soft Hubbard gap beyond the mean-field level. The formation of the soft Hubbard gap cannot be attributed to a conventional theory by Efros and Shklovskii (ES) owing the emergence of soft gaps to the long-range Coulomb interaction. Indeed, based on a picture of multivalley energy landscape, we propose a phenomenological scaling theory, which predicts a scaling of the DOS, $A$ in energy $E$ as $A(E)\propto \exp[-(-\gamma\log |E-E_\mathrm{F}|)^d]$. Here, $d$ is the spatial dimension, $E_\mathrm{F}$ is the Fermi energy and $\gamma$ is a non-universal constant. This scaling is in perfect agreement with the numerical results. We further discuss a correction of the scaling of the DOS by the long-range part of the Coulomb interaction, which modifies the scaling of Efros and Shklovskii. Furthermore, explicit formulae for the temperature dependence of the DC resistivity via variable-range hopping under the influence of the soft gaps are derived. Finally, we compare the present theory with experimental results of SrRu$_{1-x}$Ti$_x$O$_3$.

%% file: chap01.tex
\section{Introduction}
\subsection{Single-particle gaps in insulators}
\begin{figure}[h]
 \centering
 \includegraphics[width=\graphwidthsingle,clip]{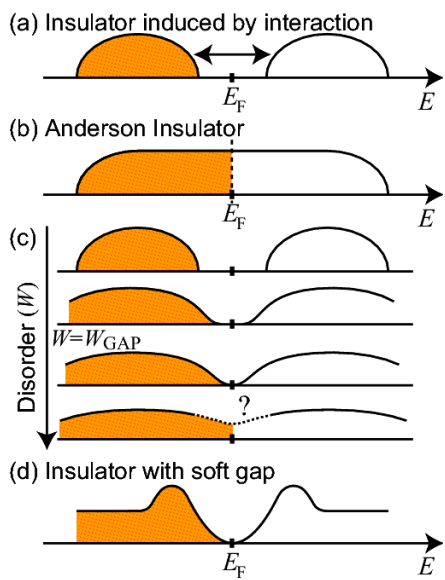}
 \caption{\coloronline Schematic of insulators with different types of gaps in the single-particle density of states: (a) Insulator induced by interaction, (b) Anderson insulator, (c) Disorder dependence of a single-particle gap and (d) Insulator with soft gap.}
 \label{fig:gap}
\end{figure}
Strongly-correlated electron systems continue to be a challenging issue of the condensed matter physics. Especially, metal-insulator transitions have been attracting much attention, because of various phases found in their vicinities~\cite{Imada98}. When the electron correlation becomes dominant compared to the kinetic energy, the ground state undergoes a transition from a metal into a correlation-induced insulator. The Mott transition at specific electron density is a typical example~\cite{mott49}. In the Mott insulator, the electrons are localized  on the individual atomic orbitals to avoid the strong on-site Coulomb repulsion, leading to the opening of a gap in the single-particle DOS as illustrated in Fig.~\ref{fig:gap} (a). There only the spin and orbital degrees of freedom remain at low energies. The absence of the single-particle excitations at low energies is a common feature of correlation-induced insulators, such as antiferromagnetic insulators and charge-ordered insulators.

Another source of the localization in the strongly-correlated electron systems is disorder (or randomness), which is inevitably present in real materials. The disorder drives the metal-insulator transition as the Anderson transition~\cite{anderson58, Abrahams79, Kramer93}. There, the insulators are characterized not by the vanishing carrier number but by a vanishing relaxation time accompanying the quantum localization of the wave functions by the impurity scattering. In contrast to the Mott insulator, the Anderson insulators without electron correlations exhibit no gap as illustrated in Fig.~\ref{fig:gap} (b), indicating that the gapless single-particle excitations are essential in determining their physical properties at low energies. This makes the Anderson insulators completely different from the Mott insulators.

Since the electron correlation and randomness inevitably coexist in real materials, clarification of the single-particle excitations under the coexistence is important for understanding their physical properties. However, since the Mott and Anderson insulators have qualitatively different low-energy excitations, the DOS at low energies under their coexistence is highly-non-trivial, which is the main topic of this paper. Let us consider how a single-particle gap behaves when fluctuating random potentials are introduced as disorder into the \textit{pure} Mott insulator. When the disorder strength is weak enough, the single-particle gap might survive despite appearance of localized impurity levels induced within the gap as illustrated in Fig.~\ref{fig:gap} (c). With the increase of disorder strength $W$, width of distribution of random potentials, however, the gap amplitude gradually decreases and finally the gap collapses at $W=W_\mathrm{GAP}$. One might think that for $W>W_\mathrm{GAP}$, the gap completely closes and the DOS at $E_\mathrm{F}$ becomes nonzero. Namely, the ground state undergoes a transition to a simple Anderson insulator, where the gapless single-particle excitations dominate the low-energy physics. Unfortunately, this naive expectation is not correct.

In their seminal work, Efros and Shklovskii~\cite{Efros-Shklovskii} (ES) considered an amorphous or a doped crystalline semiconductor, where the Coulomb interaction is nearly unscreened. They showed that assuming nonzero $A(E_\mathrm{F})$ in the ground state in the presence of the long-range Coulomb interaction, the ground state is unstable against a particle-hole excitation between localized states at $E_\mathrm{F}$, which is not a single-particle excitation but a multiple excitation from the ground state. In other words, this means that \textit{the supposed ground state} with nonzero $A(E_\mathrm{F})$ is relaxed by electron-hole excitations to reduce $A(E_\mathrm{F})$ under the influence of the long-range Coulomb interaction. As a result of this excitonic effect, \textit{soft Coulomb gap} indeed opens and the DOS at $E_\mathrm{F}$ remains zero even for $W>W_\mathrm{GAP}$ as illustrated in Fig.~\ref{fig:gap} (d), where the DOS is scaled near $E_\mathrm{F}$ as
\begin{equation}
 A(E) \propto {| E-E_\mathrm{F}|}^{d-1}. \label{eq:ES-scaling}
\end{equation}
Here $d$ is the spatial dimension. Note that although we focus on half filling in Fig.~\ref{fig:gap}, the soft Coulomb gap is not restricted to half filling.

It is important to emphasize that this power-law scaling is not restricted to the critical point which separates the correlation-induced and Anderson insulators. Instead, according to the ES theory, this power-law scaling generically dominates the whole insulating phase, which indicates that the insulator region is always critical. Indeed, the existence of the Coulomb gap was confirmed by numerical simulations~\cite{Baranovskii79, Davies82, Davies84} and in electron tunneling experiments~\cite{McMillan81, Hertel83, Massey95} later.

\subsection{Recent studies on coexistence of short-range interaction and disorder}
When the screening gets stronger \textit{e.g.,} near metal-insulator transitions with divergence of the dielectric constant, the effects of the long-range part of the Coulomb interaction are restricted to lower and lower energies. In this case, the soft Coulomb gap arising from the ES mechanism is expected to shrink to extremely low energies and the effect of the short-range part practically determines electronic structures in the experimentally accessible energy scale. However, within the ES theory, short-range interactions do not generate soft gaps, because the excitonic effect is negligible for the short-range interaction. Therefore, according to the ES theory, one could speculate that the soft gap vanishes in the insulator near the metal-insulator transition.

Although the coexistence of the interaction and randomness is common in strongly-correlated materials, particularly perovskite-type compounds with B-site substitution AB$_{1-x}$B$^\prime_x$O$_3$ offer a suitable stage for the investigation of the combined effects. SrRu$_{1-x}$Ti$_x$O$_3$ is a promising candidate for this purpose. One of the end component SrRuO$_3$ is a correlated ferromagnetic metal ($T_\mathrm{C} = 165~\mathrm{K}$); the other end component SrTiO$_3$ is a band insulator with a wide band gap ($\simeq 3.2~\mathrm{eV}$). The 4$d$ Ru band is located at the Fermi energy, while the Ti 3$d$ band is well separated from the Ru 4$d$ band around the Fermi energy. Thus the Ti atoms act as impurities for the itinerant Ru 4$d$ electrons, and the potential height of the introduced disorder is as much as of the order of 1 eV. As a consequence, around $0.3<x<0.5$, the material undergoes a metal-insulator transition into a correlated Anderson insulator~\cite{Kim05}. In the vicinity of the metal-insulator transition, indeed, recent photoemission results of SrRu$_{1-x}$Ti$_x$O$_3$~\cite{Kim06, Maiti07} indicates breakdown of the ES scaling in 3D. Although the power law with the exponent $\alpha=2$ is expected from the ES theory for 3D, the exponent obtained by the fitting of the experimental data ($\simeq 1.2$) is clearly different from $\alpha=2$. The breakdown of the ES theory has been observed in other materials such as LaNi$_{1-x}$Mn$_x$O$_3$~\cite{Sarma98}, indicating the ubiquity of unconventional soft gaps in the vicinities of the metal-insulator transitions. The deviation from the ES theory near the metal-insulator transition was also observed in a numerical study with the unrestricted Hartree-Fock approximation~\cite{Epperlein97}. 

Despite long-time efforts to go beyond the ES theory, however, effects of short-range interaction have not been fully understood yet because of the difficulties in handling quantum effects, which is important for short-range interaction. The Anderson-Hubbard model is one of the minimal models of real strongly-correlated materials under the coexistence of the interaction and randomness. The Anderson-Hubbard Hamiltonian is defined by
\begin{equation}
\mathcal{H}=-t\sum_{\langle i,j \rangle,\sigma}c_{i\sigma}^{\dagger}c_{j\sigma} + U \sum_{i} n_{i \uparrow}n_{i \downarrow}+ \sum_{i, \sigma} (V_{i}-\mu) n_{i\sigma},\!\!\label{eq:ham}
\end{equation}
with $\Ns$ sites and $\Ne$ electrons, where $t$ is a hopping integral, $U$ is the on-site repulsion, $c_{i\sigma}^\dagger$ ($c_{i\sigma}$) is the creation (annihilation) operator for an electron with spin $\sigma$ on the site $i$. The number operator is defined by $n_{i\sigma}=c_{i\sigma}^\dagger c_{i\sigma}$ and $\mu$ is the chemical potential. In addition to the usual Hubbard Hamiltonian representing the itinerancy and the short-range interaction (namely, the first and the second terms of \Eq~(\ref{eq:ham})), disorder is represented by the spatially uncorrelated spin-independent random potential $V_i$.

Many numerical techniques have been applied to the Anderson-Hubbard model. For example, the quantum Mote Carlo (QMC) method was applied in 1D~\cite{Sandvik94, Otsuka98}, 2D~\cite{Ulmke97} and 3D~\cite{Otsuka00}. Furthermore, the dynamical mean-field theory was extended to disordered systems~\cite{Dobrosavljevic97}. At infinite dimensions, Dobrosavljevi\'{c} and Kotliar formulated a variant of the dynamical mean-field theory (DMFT)~\cite{Dobrosavljevic97, Dobrosavljevic98}, so-called statistical DMFT, which is exact in the limit of the infinite coordination number or in the non-interacting limit. By using the site-dependent bath functions, statistical DMFT can partially treat spatial correlations. Subsequently, Dobrosavljevi\'{c} \textit{et al.} derived a mean-field theory~\cite{Dobrosavljevic03}, which is simpler but ignores the spatial correlations. Another approach is the (unrestricted) site-dependent Hartree-Fock (HF) approximation. Although it treats the electron correlation in the mean-field level, it is certainly beyond the mean-field theory of the type that allows only spatially uniform mean fields. In fact, it can describe the inhomogeneity of the electronic structures by using site-dependent mean-fields. Tusch \textit{et al.} obtained the ground-state phase diagram including both of the magnetic and charge degrees of freedom in 3D~\cite{Tusch93}. 

Recently, several numerical studies have reported suppression of the DOS at the Fermi energy even for short-range interaction. For example, soft gaps were reported in a HF study in 3D~\cite{Fazileh06}. Although they claimed a power-law scaling of the DOS with $\alpha \simeq 0.5$, the origin of the soft gap has not been clarified at all. Recent numerical studies in 2D by using the exact diagonalization showed a dip of the DOS near $E_\mathrm{F}$~\cite{Chiesa08}. These strongly suggest the presence of an unconventional mechanism which suppresses the DOS even with short-range interaction. However, a numerical study with statistical DMFT~\cite{Dobrosavljevic97} claimed nonzero $A(E_\mathrm{F})$ in the insulating phase, and even the divergence of $A(E_\mathrm{F})$ toward the metal-insulator transition from the metallic side, which completely disagrees with the HF results. On the other hand, in several mean-field studies which ignores the spatial correlations, no singularity was reported in the DOS~\cite{Dobrosavljevic03, Byczuk05}. We clearly need further studies for comprehensive understanding of the short-range case.

In this paper, through numerical analyses of the 3D Anderson-Hubbard model, we show that there exists a soft gap even though only short-range interaction is present in the Anderson-Hubbard model. We call this unconventional soft gap \textit{soft Hubbard gap}. We show numerical evidences of the soft Hubbard gaps within the HF approximation in 1D, 2D and 3D. Further support by the exact diagonalization in 1D is given. In order to clarify the origin of the soft Hubbard gap, we propose a phenomenological theory based on a picture of \textit{multivalley energy landscape}, which corresponds to emergence of many excited states degenerated with the ground state. Our scaling theory predicts an unconventional scaling of the density of states $A(E)$ in energy $E$ as $A(E)\propto \exp[-(-\gamma\log |E-E_\mathrm{F}|)^d]$. Here $\gamma$ is a non-universal constant. We show that this predicted scaling is consistent with the numerically-observed scaling. A part of the numerical evidences for the soft Hubbard gap and the scaling theory have already been given in a letter briefly~\cite{Shinaoka08}. In this paper, however, we analyze the ground-state phase diagram of the 3D Anderson-Hubbard model in greater detail. Especially, the criticality of the metal-insulator transitions is clarified from a viewpoint of the formation of the soft Hubbard gap. Furthermore, by extending our scaling theory, we clarify effects of the long-range part of the Coulomb interaction responsible for low-energy excitations. Especially, we show that the ES theory is seriously modified in the presence of the long-range Coulomb interaction, when we consider the multiply-excited states by extending our scaling theory originally constructed for the short-range interaction. In order to inspire future experimental efforts to examine the validity of the present fundamental proposal, we study temperature dependence of the DC resistivity in the presence of the soft gap. We further compare the experimental results for SrRu$_{1-x}$Ti$_x$O$_3$ with the present theory.

This paper is organized as follows; In {\S}~\ref{SEC:MODEL-METHOD}, we introduce the Anderson-Hubbard model and numerical methods employed. In {\S}~\ref{SEC:NUMERICAL}, we show numerical results of the Anderson-Hubbard model in 3D and 1D within the Hartree-Fock approximation as well as by the exact diagonalization. Section \ref{SEC:THEORY} is devoted to the scaling theory of the soft gap and its extension to discrete distribution functions of random potentials and the long-range Coulomb interaction. In {\S}~\ref{sec:transport}, we derive transport properties in the presence of the soft gap. Comparisons with experimental results are also given. The summary and discussion are given in {\S}~\ref{SEC:CONCLUSION}.

%% file: method.tex
\section{Model and Method}\label{SEC:MODEL-METHOD}
\subsection{Anderson-Hubbard Model}
In this paper, we analyze the Anderson-Hubbard model, whose Hamiltonian is defined by \Eq~(\ref{eq:ham}). We employ a cubic lattice for $d=3$, a square lattice for $d=2$ and a chain lattice for $d=1$. We take the lattice spacing as the length unit. The spin-independent random potential $V_i$ representing randomness is assumed to follow two models of the distribution $P_V(V_i)$: the box type of width $2W$, 
\begin{eqnarray}
P_V(V_i)&=&
\left\{
	\begin{array}{ll}
		\frac{1}{2W} & (|V_i|<W) \\
		0 & (\mbox{otherwise})
	\end{array}
\right.
\end{eqnarray}
with the average $\langle V_i \rangle =0$, and the Gaussian type, 
\begin{eqnarray}
	P_V(V_i)&=&\frac{1}{\sqrt{2\pi} \sigma} \exp\left( -\frac{V_i^2}{2\sigma^2}\right), \label{eq:Gaussian}
\end{eqnarray}
where $\sigma^2 = W^2/12$ with the average $\langle V_i \rangle =0$. Because $W$ is proportional to the standard deviation of the distribution for the both models, $W$ is the parameter to control the strength of disorder. For both the two distributions, $\mu = U/2$ corresponds to half filling. 

\subsection{Hartree-Fock approximation}
We first employ the Hartree-Fock (HF) approximation, where the wave function is approximated by a single Slater determinant consisting of a set of orthonormal single-particle orbitals $\{ \phi_n \}$ ($n$ is an orbital index). Within this trial wave function, the variational principle leads to the HF equation as  
\begin{equation}
\{\mathcal{H}_0+U \sum_{i} ( \langle n_{i \downarrow} \rangle n_{i \uparrow}+ \langle n_{i \uparrow} \rangle n_{i \downarrow} ) \} \phi_n= \epsilon_n \phi_n,
\end{equation}
where $\mathcal{H}_0$ is the one-body part of the Hamiltonian and we neglect $\langle c_{i \uparrow}^\dagger c_{i \downarrow}\rangle$. Here $\langle n_{i\sigma}\rangle$ are the site-dependent mean fields. Later, we will show that the inhomogeneity of the electronic structures is necessary for the formation of the soft Hubbard gap. To find a site-dependent mean-field solution $\langle n_{i \sigma} \rangle$ for the HF equations, we employ the iterative scheme starting from an appropriate initial guess until the convergence condition, $|\Delta n_{i\sigma}|<10^{-5}$ ($\forall i\sigma$) is satisfied. Here $\Delta n_{i\sigma}$ denotes a change in the mean field at the site $i$ with spin $\sigma$, $n_{i\sigma}$ before and after an iteration. Initial guesses of mean fields employed in our calculations will be described later in each case. In order to accelerate the convergence of the iteration, we employ the Anderson mixing~\cite{Anderson65}, which is easily applicable to the 3D case because its computational cost scales as $O(\Ns)$ and it is not heavy. 

In general, many stable mean-field solutions may coexist for a given realization of the random potentials. Therefore, after repeating the calculations for several different initial guesses of the mean fields as the ground state, we employ the solution that has the lowest energy.

\subsection{Exact diagonalization}
In this paper, we further employ the exact diagonalization method in 1D. Because the exact diagonalization method takes into account quantum fluctuations ignored by the Hartree-Fock approximation, it is suitable for the examination of robustness of our Hartree-Fock results against the quantum fluctuations. By diagonalizing the full Hamiltonian matrix by using LAPACK routines~\cite{LAPACK}, we obtain all the eigenstates. Then the single-particle density of states is calculated by 
\begin{equation}
A(E)=
\left\{
	\begin{array}{l}
		\sum_{n}\sum_{i\sigma} {|\langle n;\Ne+1 | c_{i\sigma}^\dagger|0;\Ne\rangle|}^2 \\
		\ \ \ \ \ \ \ \ \times\delta(E-E_+(n)) \ \ \ \ (E>E_\mathrm{F}) \\
		\sum_{n}\sum_{i\sigma} {|\langle n;\Ne-1 | c_{i\sigma}^\dagger|0;\Ne\rangle|}^2 \\
    \ \ \ \ \ \ \ \ \times\delta(E-E_-(n)) \ \ \ \ (E<E_\mathrm{F}),
	\end{array}
\right.
\end{equation}
where $\Ne$ is the number of electrons in the ground state, $|n;N \rangle$ is the $n$-th eigenstates with $N$ electrons ($n\ge 0$), and $|0;\Ne\rangle$ is the ground state. The single-particle excitation energies $E_\pm(n)$ are defined as
\begin{equation}
	E_\pm(n) = E-E_{n}(\Ne\pm1)+E_{0}(\Ne)-E_\mathrm{F}.
\end{equation}
Here $E_n(N)$ is the energy of the $n$-th eigenstates with $N$ electrons, $|0;\Ne\rangle$. In order to reduce the computational cost, we divide the full Hilbert space into subspaces with a fixed set of $(N_\uparrow, N_\downarrow)$, where $N_\uparrow$ and $N_\downarrow$ denote the number of up-spin electrons and that of down-spin electrons, respectively. The maximum dimension of the subspaces is 400 ($\Ns=6$).

%% file: numerical.tex
\section{Numerical Results}\label{SEC:NUMERICAL}
In this section, we analyze the Anderson-Hubbard model in 1D and 3D numerically.

Our main new results are the following:
\begin{enumerate}
 \item Determination of the ground-state phase diagram of the Anderson-Hubbard model in 3D within the HF approximation,
 \item Discovery of the unconventional soft gap driven by the short-range interaction regardless of electron filling and spatial dimensionality,
 \item Clarification of the unconventional scaling of the soft gap; the DOS decays faster than any power law in 3D, while it follows a power law in 1D,
	 \item Clarification of the criticality of the metal-insulator transitions.
\end{enumerate}
We first analyze the Anderson-Hubbard model in 3D within the HF approximation. Further numerical evidences of the soft gap in 1D are given within the Hartree-Fock approximation. We also show exact-diagonalization results in 1D beyond the mean-field level. Detailed scaling analyses within the Hartree-Fock approximation in 2D will be given in the next section \S~\ref{sec:scaling}. In the following, we focus on half filling unless otherwise stated. Although a part of the results have already been reported briefly~\cite{Shinaoka08}, in this paper, we show more detailed numerical analyses which were omitted in the previous letter, \textit{e.g.} the criticality of the metal-insulator transitions.

\subsection{Three dimensions}
\subsubsection{Ground-state phase diagram}
\begin{figure}
 \centering
 \includegraphics[width=0.425\textwidth,clip]{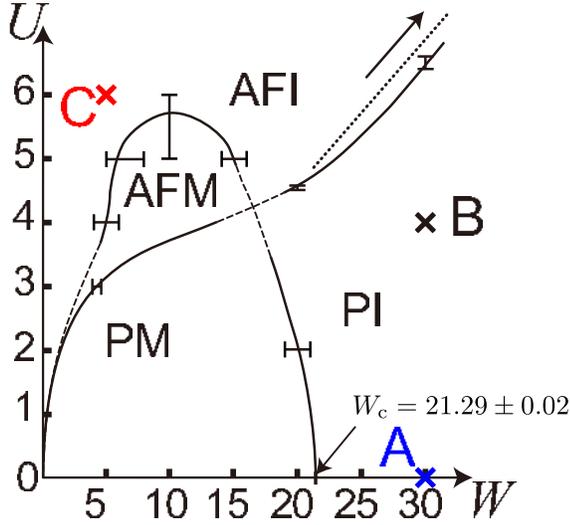}
 \caption{\coloronline Ground-state phase diagram of 3D Anderson-Hubbard model at half filling for Gaussian distribution of $P_V$. Abbreviations are: AFI, antiferromagnetic insulator; AFM, antiferromagnetic metal; PI, paramagnetic insulator (Anderson insulator); PM, paramagnetic metal. The dotted line denotes the asymptotic scaling of the antiferromagnetic transition for $U\gg t$ and $W\gg t$. The critical value of the metal-insulator transition in the non-interacting limit is given by $W_\mathrm{c}=21.29\pm 0.02$.~\cite{Slevin99}}
 \label{fig:pd}
\end{figure}

We study the ground state of the 3D Anderson-Hubbard model with the Gaussian distribution of $P_V$ in detail. A part of the results have already been reported briefly~\cite{Shinaoka08}. In our 3D study, we take the hopping integral $t$ as the energy unit. Figure~\ref{fig:pd} shows the calculated ground-state phase diagram within the HF approximation. At $U=0$, the Anderson-Hubbard model undergoes a metal-insulator transition (Anderson transition) from the paramagnetic metal (PM) to the paramagnetic insulator (PI) at a finite strength of the disorder, $W_\mathrm{c}=21.29\pm0.02$~\cite{Slevin99}. On the other hand, at $W=0$, since the system is half-filled with a perfect nesting, the ground state is the antiferromagnetic insulator (AFI) for any nonzero value of $U$. Here, we discuss the ground-state phase diagram for $U, W>0$. First, we focus on the spin degree of freedom. For $W>0$, the ground state is paramagnetic near $U=0$. With increasing the interaction, the ground states undergoes an antiferromagnetic transition at a critical point $U_\mathrm{c}~(>0)$. Within the resolution of our calculation, $U_\mathrm{c}$ monotonically increases as the disorder strength $W$ increases. Next, we focus on the charge degree of freedom. The ground state is insulating for $U, W\gg1$, which contains AFI as well as paramagnetic insulator (PI) (PI is usually identified as Anderson insulator). Metallic phases are restricted to a dome-like region ($U<6$ and $W<25$). In addition to a paramagnetic metal (PM), we found an antiferromagnetic metal (AFM). In the following, we will show detailed analyses of the ground-state phase diagram. 

\subsubsection{Antiferromagnetic transitions}
\begin{figure}
 \centering
 \includegraphics[width=0.425\textwidth,clip]{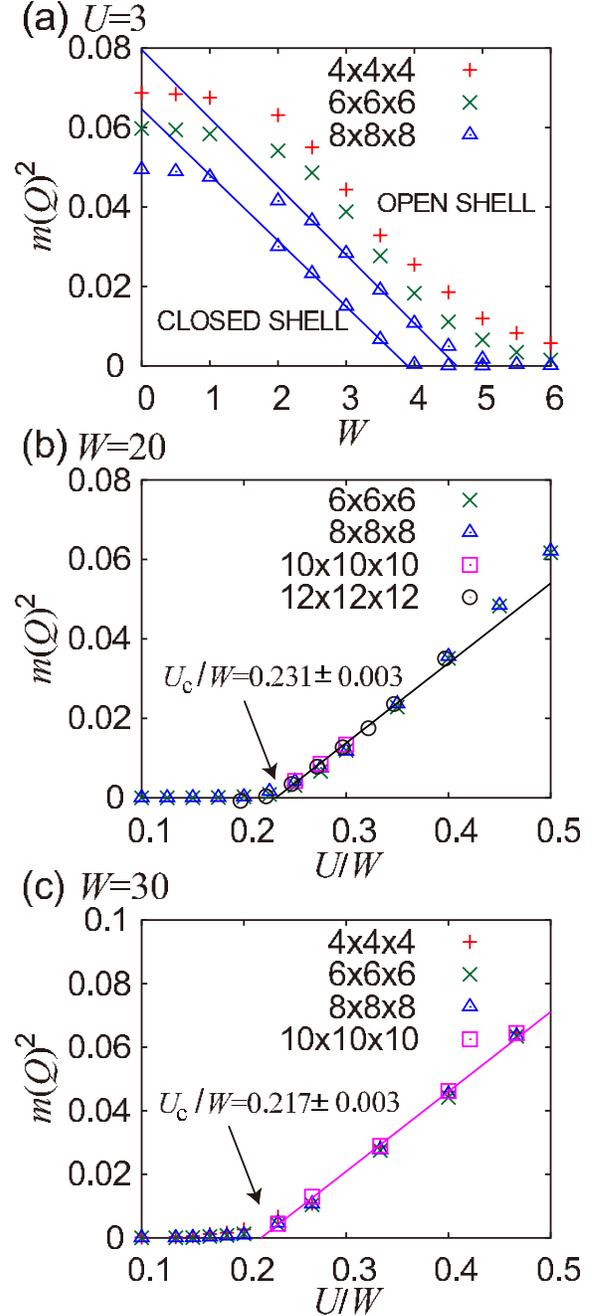}
 \caption{\coloronline Square of antiferromagnetic order parameter $m(Q)^2$ at $U=3$, $W=20$ and $W=30$ for (a), (b) and (c), respectively. The solid lines are the fit by the mean-field criticality $\beta=1/2$. Results for $4\times 4\times 4$ and $6\times 6\times 6$ are shown only for the open-shell boundary condition. In (b) and (c), we only show results from the open-shell boundary condition.}
 \label{fig:mQ}
\end{figure}
We first discuss the antiferromagnetic transition. The antiferromagnetic order parameter $m(Q)$ is defined by 
\begin{eqnarray}
 m(Q) &=& \frac{1}{\Ns} \left|\sum_i \langle S_i^z \rangle e^{i Q r_i}\right|, 
\end{eqnarray}
where $Q=(\pi, \pi, \pi)$. At $W=0$, since the system is half-filled with the perfect nesting condition satisfied, the antiferromagnetic order vanishes in an essentially-singular way toward $U=0$ as~\cite{Hirsch85}
\begin{eqnarray}
m(Q) = \frac{t}{U}\exp\left(-\frac{2\pi t}{U}\right).
\end{eqnarray}
While the amplitude of the antiferromagnetic gap given by $Um(Q)$ also follows essentially-singular scaling and is small near $U=0$, $U_\mathrm{c}$ increases sharply from zero with the increasing disorder $W$.

We consider the antiferromagnetic transition in the atomic limit ($U, W \gg t$). In this limit, there are two types of sites: singly occupied sites for $|V_i|<U/2$, and doubly occupied or empty sites for $V_i<-U/2$ or $V_i>U/2$, respectively. Thus the Anderson-Hubbard model reduces to a site-diluted antiferromagnet~\cite{Heidarian04}, where the antiferromagnetic transition coincides with percolation of the singly occupied sites within the mean-field level, because of the antiferromagnetic interaction between neighboring singly occupied sites. Thus $U_\mathrm{c}$ is determined by 
\begin{equation}
	\int_{-U_\mathrm{c}}^{+U_\mathrm{c}} P(V_i) dV_i  = p_c,
\end{equation}
where $p_c \simeq 0.311608$ is the site-percolation threshold of the cubic lattice. By solving this equation for the Gaussian distribution \Eq~(\ref{eq:Gaussian}), one obtains $U_\mathrm{c}/W \simeq 0.2315$ in the atomic limit. This asymptotic scaling is plotted as the dotted line in Fig.~\ref{fig:pd}. As we will show below, the phase boundary numerically obtained starts following this scaling with increasing $W$ and $U$.

In order to determine the antiferromagnetic transition line in the intermediate region, we calculate the antiferromagnetic order parameter with cubic unit cells of $L\times L\times L$. We employ two boundary conditions: closed-shell and open-shell boundary conditions. At $U=0$ and $W=0$, the periodic boundary condition corresponds to open-shell configurations for $L=4n$ and closed-shell configurations for $L=4n+2$, where $n$ is a positive integer. In contrast, the antiperiodic boundary condition in 3D corresponds to closed-shell configurations for $L=4n$ and open-shell configurations for $L=4n+2$. Therefore we define the open-shell boundary condition as the periodic boundary condition for $L=4n$ and the antiperiodic boundary condition for $L=4n+2$. The closed-shell boundary condition is defined as the opposite cases. For finite-size lattices, the antiferromagnetic order parameter calculated with the open-shell boundary condition is larger than that with the closed-shell boundary condition, because of an enhancement of the electron correlation for the open-shell boundary condition.

We employ four kinds of initial guesses of mean fields:
\begin{enumerate}
 \item Uniform and antiferromagnetically-ordered states,
 \item Ground states in the limit of $t\rightarrow 0$ degenerating with each other within the second-order perturbation with respect to $t$, which consists of locally-antiferromagnetically-ordered clusters parted by doublons and holons,
 \item Paramagnetic states with uniform charge distribution,
 \item States with random charge-spin distributions.
\end{enumerate}
All these initial guesses are further perturbed by different choices of random noise in both of the spin and charge sectors before putting into the self-consistent iteration. We typically need several tens of initial guesses for convergence of the antiferromagnetic order parameters. 
Figure~\ref{fig:mQ} (a) shows the calculated antiferromagnetic order parameter at $U=3$ with the open-shell and closed-shell boundary conditions, respectively. Error bars smaller than the symbols are dropped for the simplicity of the figure. For each boundary condition, as the system size increases, the antiferromagnetic order parameter well converges to the mean-field critical behavior of $\beta=1/2$. We estimate the upper and lower limits of the critical point by fitting of the data with the open-shell boundary condition and those with the closed-shell boundary condition, respectively. 

At $W/t=20, 30$, we show the antiferromagnetic order parameter calculated only with the open-shell boundary condition, because difference in the calculated antiferromagnetic order parameter was found to be negligible between the two boundary conditions. As shown in Fig.~\ref{fig:mQ} (b) and (c), the antiferromagnetic order parameter again well converges to the mean-field critical behavior, as the system size becomes larger. The estimated antiferromagnetic transition points of $U_\mathrm{c}/W=0.231\pm0.003, 0.217\pm0.003$ for $W/t=20, 30$ are close to the atomic-limit value ($U_\mathrm{c}/W \simeq 0.2315$), establishing the validity of our calculations. The error bars are determined by the fitting of the data with the system size $12\times 12\times 12$ for $W=20$, $10\times 10\times 10$ for $W=30$, respectively. This correct asymptotic behavior of $U_\mathrm{c}$ was not reproduced in the previous HF study~\cite{Tusch93}.

\begin{figure}
 \centering
 \includegraphics[width=0.425\textwidth,clip]{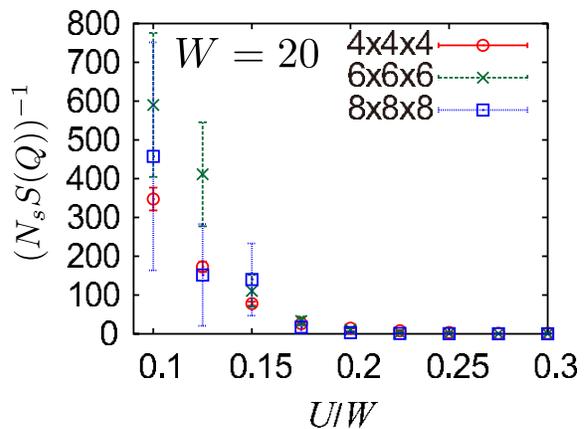}
 \caption{\coloronline Inverse of spin structure factor $S(Q)$ multiplied by the number of sites $\Ns$ as a function of $U$ at $W=20$.}
 \label{fig:SQ}
\end{figure}

On the other hand, K. Byczuk \textit{et al.} found a non-monotonic behavior of $U_\mathrm{c}$ for the antiferromagnetic transition as a function of the increasing disorder strength $W$ within DMFT with spin degrees of freedom~\cite{Byczuk08}. Namely, they found that $U_\mathrm{c}$ becomes zero as a function of $W$ at the Anderson transition point in the non-interacting limit. In contrast, we found a monotonic increase of $U_\mathrm{c}$ in our 3D study, which we believe more reasonable. We show a further numerical evidence at $W=20$. We define spin structure factor $S(q)$ as 
\begin{eqnarray}
 S(q)=\frac{1}{\Ns}\left|\sum_{i,j} \langle S_i \cdot S_j \rangle e^{iq\cdot (r_i-r_j)} \right|.
\end{eqnarray}
In Fig.~\ref{fig:SQ}, we plot spin structure factor at $Q=(\pi,~\pi,~\pi)$ for $W=20$. The quantity $\Ns S(Q)$ diverges toward the antiferromagnetic transition point approaching from the limit of $U=0$, which further supports the existence of the antiferromagnetic transition around $U/t \simeq 4$.

\subsubsection{Metal-insulator transitions}\label{SEC:MIT}
Next, we discuss the metal-insulator transitions. We identify insulating phases by extrapolation of the localization lengths $\xi$ to the bulk limit. The localization length $\xi$ is defined by the asymptotic behavior of single-particle orbitals near $E_\mathrm{F}$ at long distances as
\begin{eqnarray}
	\phi_n &\propto& \exp(-r/\xi), \label{eq:localization-length}
\end{eqnarray}
where $r$ is the distance from the center of the orbital. In order to observe the asymptotic behavior easily, we employ pseudo-1D unit cells of $L\times L\times M$, where $M \gg L$ as has been employed in Refs.~\citen{Pichard81a, Pichard81b, MacKinnon81}. In fact, we adopt $M=1000, 250, 250, 100$ for $L=4, 6, 8, 10$, respectively. We use single-particle orbitals within an energy window of the width $0.01$ around the Fermi energy, namely, $|E-E_\mathrm{F}|<0.01$ to calculate the localization lengths.

\begin{figure}[h]
 \centering
 \includegraphics[width=0.425\textwidth,clip]{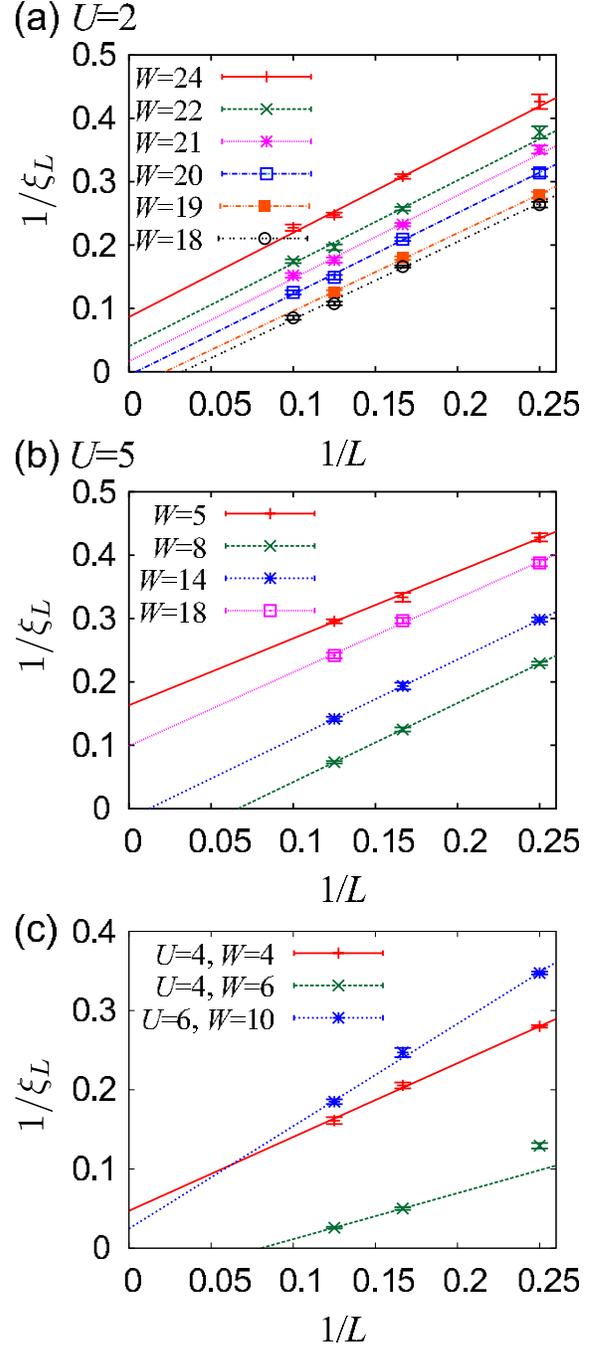}
 \caption{\coloronline Quasi-1D localization lengths as a function of inverse length $1/L$.}
 \label{fig:localization_length}
\end{figure}

Figure~\ref{fig:localization_length} shows the extrapolation of the calculated localization lengths for $U=2, 4, 5, 6$ to the bulk limit. In insulating phases, localization lengths are extrapolated to finite values. For $U<6$ and with increasing $W$ from $0$, metals appear from AFI as in the 2D result~\cite{Heidarian04}, with further reentrant transition to insulators (AFI or PI) at larger $W$. Even in antiferromagnetically-ordered phases, there exist metal-like regions, where the antiferromagnetic order parameter is locally reduced due to relatively large fluctuations of $V_i$. When $W$ is small enough, the system remains insulating even though the DOS is gapless, because these metal-like regions are small in size and isolated from each other. With increasing $W$, however, they grow in size and finally percolates, corresponding to the first metal-insulator transition into metals (AFM or PM). Further increase of the disorder \textit{i.e.}, $W \gg t$, results in the reentranst metal-insulator transition into insulators (AFI or PI). It should be noted that because the Gaussian-type distribution is unbounded, the gapless regions appear and the antiferromagnetic gap closes as soon as the disorder strength $W$ becomes nonzero. This indicates $W_\mathrm{GAP}=0$. Thus there is no real gap for $W>0$ in the case of the Gaussian-type distribution of $P_V$. However, this is not a generic feature for $P_V$ bounded in a finite region.

\begin{figure}[h]
 \centering
 \includegraphics[width=0.425\textwidth,clip]{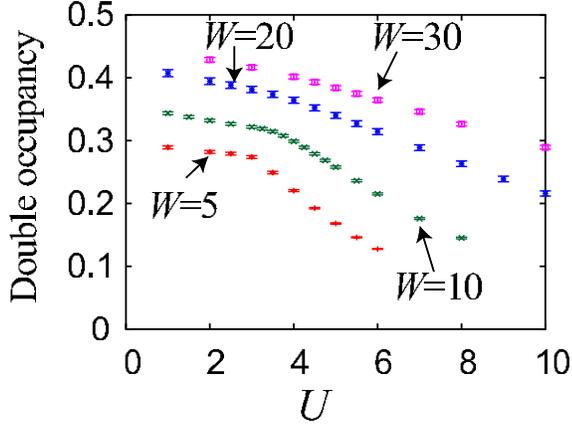}
 \caption{\coloronline Double occupancy with the open-shell boundary condition as functions of $U$. The system sizes are $8\times 8\times 8$ for $W=5$ and $20$, $10\times 10\times 10$ for the rest. The double occupancy increases as $W$ increases for a fixed value of $U$. This is consistent with the fact that the double occupancy has its maximum value $0.5$ in the limit $W=+\infty$, where only an equal number of doubly occupied sites and empty sites exist in the ground state. It should be mentioned that the double occupancy is $0.25$ at $U=0$ and $W=0$, and $0$ in the limit $U=+\infty$.}
 \label{fig:dd-U}
\end{figure}

\begin{figure}[h]
 \centering
 \includegraphics[width=0.425\textwidth,clip]{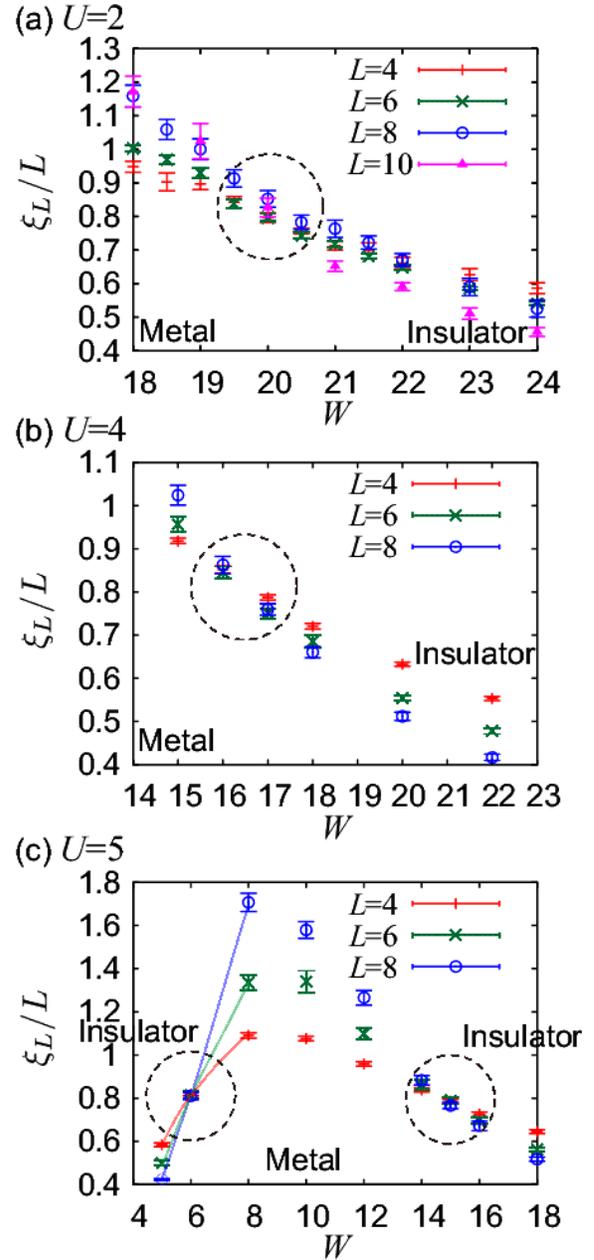}
 \caption{\coloronline Finite-size scaling plot of localization lengths at $U=2$, $U=4$ and $U=5$. The metal-insulator transition points are denoted by the centers of the circles.}
 \label{fig:scaling_localization_length}
\end{figure}

In the previous DMFT study, K. Byczuk \textit{et al.} found a first-order transition between the Mott insulator and the metal~\cite{Byczuk05}. In contrast, the double occupancy in our 3D study exhibits a jump neither at the metal-insulator transitions nor at the antiferromagnetic transitions as shown in Fig.~\ref{fig:dd-U}, indicating that they are not of the first-order but continuous. Now, we discuss the criticality of the metal-insulator transitions. In the non-interacting limit, the normalized localization lengths $\Lambda_L\equiv \xi_L/L$ for the different system sizes cross at the Anderson transition. Namely, $\Lambda_L$ is constant with respect to the width of the 1D bar $L$ when $L$ is large enough. Furthermore, the critical parameter $\Lambda_\mathrm{c}=\Lambda_L(W_\mathrm{c})$ was reported to be universal, not depending on the distribution of the random potentials~\cite{Slevin99}. In Fig.~\ref{fig:scaling_localization_length}, we show scaling plots of the localization lengths at $U=2$, $U=4$ and $U=5$. Even for the interacting cases, the normalized localization lengths indeed seem to cross at the metal-insulator transitions with a universal critical parameter $\Lambda_\mathrm{c}\simeq 0.8$. This value is clearly different from that of the non-interacting limit ($\Lambda_\mathrm{c}=0.576\pm 0.002$~\cite{Slevin99}), indicating that the metal-insulator transitions for $U>0$ belong to a universality class different from that of the Anderson transition at $U=0$. Further support for this unconventional criticality will be given in \S~\ref{sec:softHubbardgap} in the context of the density of states.

\begin{figure}
 \centering
 \includegraphics[width=0.425\textwidth,clip]{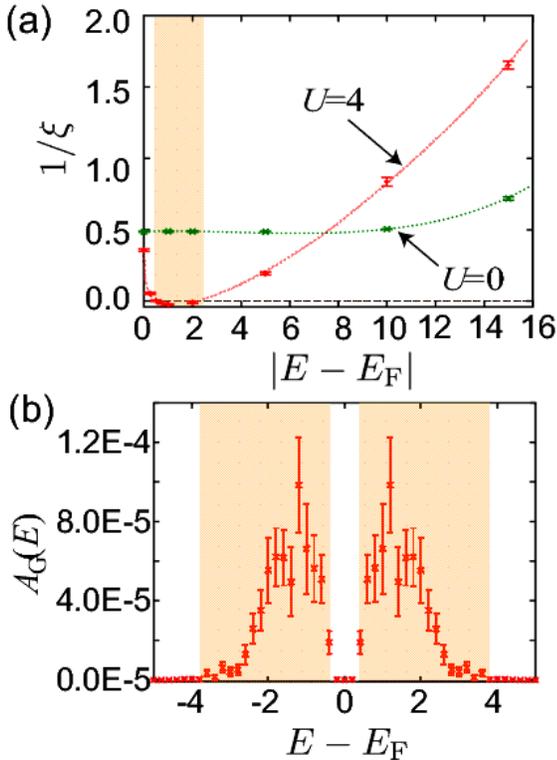}
 \caption{\coloronline (a) Energy dependence of the localization length extrapolated to the bulk limit, $\xi$ at $t=0$, $U=4$ and $W=30$. The shade denotes the energy region where the states are extended ($1/\xi = 0$). The dotted lines are guides for the eyes. (b) Geometrically-averaged DOS extrapolated to the limit of $\eta=+0$ and the bulk limit. The shade denotes the energy region where the states are extended ($A_\mathrm{G}=0$)}
 \label{fig:energy-dependent-xi}
\end{figure}

Before moving on the next section, we analyze the energy dependence of the localization length in the paramagnetic insulator (PI). In Fig.~\ref{fig:energy-dependent-xi} (a), we show the localization length extrapolated to the bulk limit, $\xi$ at $t=1$, $U=4$ and $W=30$ as well as at $t=1$, $U=4$ and $W=30$. Although the localization length is nearly independent of the energy near the Fermi energy at $U=0$, it depends on the energy in a more complex manner at $U=4$. Around the Fermi energy, namely $|E-E_\mathrm{F}|<0.5$, the inverse localization length $1/\xi$ is extrapolated to a finite value, being consistent with the fact that the ground state is insulating. Also for $|E-E_\mathrm{F}|>5$, the extrapolated localization length is finite and increases toward high energies. In the intermediate region, namely $0.5 < |E-E_\mathrm{F}| < 5$, however, $1/\xi$ is extrapolated to zero, indicating the existence of extended states. This reentrant \textit{transition} in the immediate vicinity of the Fermi energy is clearly beyond the conventional picture of the Anderson transitions without electron correlations, where there is no reentrant transition with respect to the energy~\cite{Bulka87}.

In order to reinforce the results obtained by the analyses of the localization lengths, we further analyze another order parameter for the Anderson transition, the geometrically-averaged DOS, The geometrically-averaged DOS, $A_\mathrm{G}(E)$ is defined as
\begin{eqnarray}
	A_\mathrm{G}(E) &=& \exp\left(\frac{1}{N_s}\sum_{i=1}^{N_s} \log(A_i(E))\right),
\end{eqnarray}
where the local DOS at the site $i$ is given by
\begin{eqnarray}
 A(E) &=& -\frac{1}{\pi} \mathrm{Im} G_{i,i}(E+i \eta),
\end{eqnarray}
which is exact in the limit of $\eta=+0$. Here $\eta$ ($>0$) is a broadening parameter. In the bulk limit, the geometrically-averaged DOS at the Fermi energy is another order parameter for the Anderson transitions, being nonzero when states at the Fermi energy are extended and zero when they become localized. This is because when wave functions at a given energy become localized in real space, the local DOS at that energy becomes discrete: localized wave functions at the same energy are well separated in position so the local DOS at that energy goes to zero far from the wave function centers. In Fig.~\ref{fig:energy-dependent-xi} (b), we show the geometrically-averaged DOS extrapolated to the bulk limit and further to the limit of $\eta=+0$. At low but nonzero energies, the geometrically-averaged DOS is extrapolated to a nonzero value, indicating the existence of extended states. This is consistent with the results obtained by the extrapolation of the localization length. We will discuss the implication of this unusual feature in the next subsection.

\subsubsection{Soft Hubbard gaps}\label{sec:softHubbardgap}
\begin{figure}
 \centering
 \includegraphics[width=0.425\textwidth,clip]{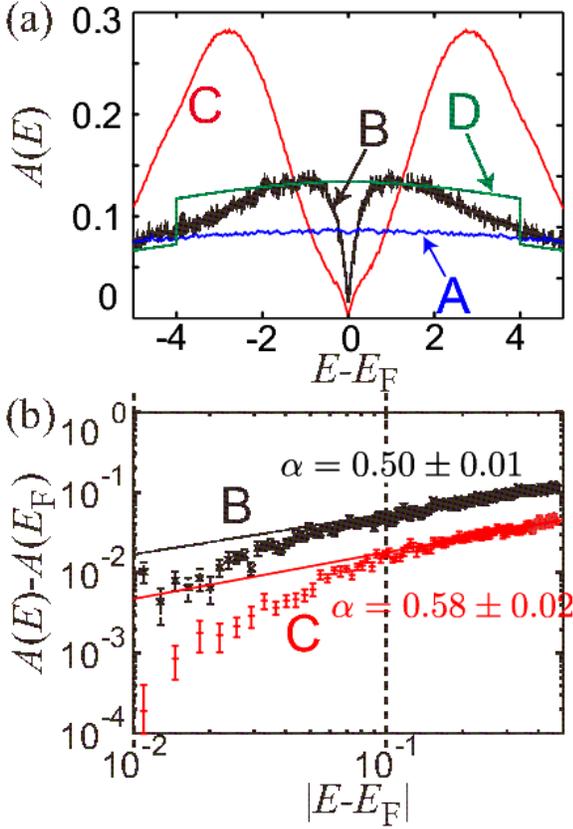}
 \caption{\coloronline DOS with system size $8\times8\times250$: A ($t=1, U=0,W=30$), B ($t=1, U=4,W=30$), C ($t=1, U=6,W=5$), D ($t=0, U=4,W=30$). (a) Linear plot. (b) Double logarithmic plot. The solid lines denote the fitting lines by a power law.}
 \label{fig:DOS}
\end{figure}

Next, we discuss the DOS. Figure~\ref{fig:DOS}(a) shows the DOS for typical parameters. Naively one might expect $A(E_\mathrm{F})>0$ for $W>0$. Indeed, there exists no soft gap in the DOS at the parameter A depicted in Fig.~\ref{fig:pd} which corresponds to the non-interacting Anderson insulator. The DOS $A(E_\mathrm{F})$ is also nonzero for $t=0$ as is seen in the case D in Fig.~\ref{fig:pd}. However, we find a soft gap over the entire insulating phases in the case of $U>0$ regardless of the antiferromagnetic order. Figure~\ref{fig:DOS}(a) shows the DOS for the typical parameters B and C, respectively. These parameters are chosen to be located in the insulating phases far from the metal-insulator transition, because the energy scale of the soft gap decreases toward the metal-insulator transition, which makes numerical analyses of the DOS difficult. A similar soft gap is seen in lower dimensions. Actually, one- and two-dimensional data will be analyzed in \S~\ref{sec:1D} and \S~\ref{sec:scaling}, respectively. We call this unconventional soft gap \textit{soft Hubbard gap}, since it is driven by the short-range interaction $U$. The significance of the soft Hubbard gap is clear because the soft gap is established irrespective of the spatial dimension and electron filling. It should be mentioned that the soft Hubbard gap is not restricted to half filling as well, as we will see in 1D. Although the soft gap is observed generically, its formation certainly has to satisfy minimal requirement: Not only the coexistence of the interaction and randomness but also the itinerancy is required for their formation. Indeed, the soft Hubbard gap vanishes at the parameter D by switching off the transfer. All of the three terms in the Anderson-Hubbard Hamiltonian (\Eq~(\ref{eq:ham})), namely itinerancy, interaction and randomness are imperative for the formation of the soft gap.

This soft gap moreover bears generically an unconventional nature: Although a power law scaling $A(E)\propto |E-E_\mathrm{F}|^{\alpha}$ with exponents $0.5<\alpha<1$ looks fit in the range $|E-E_\mathrm{F}| > 0.1$ being consistent with the previous HF study~\cite{Fazileh06}, a closer look for $|E-E_\mathrm{F}|<0.1$ shows clear deviation from the power-law scaling, and even faster decay of the DOS near $E_\mathrm{F}$ as shown in Fig.~\ref{fig:DOS} (b). The deviation from a simple power law strongly suggests that the soft Hubbard gap originates from a novel mechanism.

The formation of the soft Hubbard gap in the insulating side further supports an unconventional universality class of the metal-insulator transitions for the interacting case $U>0$ (see \S~\ref{SEC:MIT}). The critical exponent of the density of states $\beta$ is defined by
\begin{eqnarray}
	A(E_\mathrm{F})&=& |W-W_\mathrm{c}|^\beta~\mbox{(Metallic side)}.
\end{eqnarray}
For $U=0$, the critical exponent $\beta$ is zero because the density of states remains nonzero in the Anderson insulator. However, the critical exponent $\beta$ should be nonzero for $U>0$ because of the formation of the soft Hubbard gap. This further supports that the metal-insulator transition for $U>0$ belongs to a universality class different from that for $U=0$. The formation of the soft Hubbard gap in the insulating phases may be responsible for the increase of the critical parameter $\Lambda_\mathrm{c}$ from the non-interacting value ($\Lambda_\mathrm{c}=0.576\pm 0.002$) to $\simeq 0.8$ by switching on the electron correlation. Now we consider how localized states at the Fermi energy become extended toward the metal-insulator transition by approaching from the strongly-localized limit. In the presence of the soft Hubbard gap, the mean distance between localized states near the Fermi energy for $U>0$ is larger than that at $U=0$. This prevents the localized states from being hybridized with each other and percolating. Thus the single-particle wave functions at the Fermi energy remain localized in the bulk limit for $U>0$, even if the quantity $\Lambda_L=\xi_L/L$ has reached the non-interacting critical parameter $\Lambda_\mathrm{c}$.

Furthermore, the formation of the soft Hubbard is responsible for the low-lying extended excited states in the insulating phases (see Fig.~\ref{fig:energy-dependent-xi}). Because of the formation of the soft Hubbard gap at the Fermi energy, and the weight excluded and transferred from the low-energy region around the Fermi energy, the DOS forms a peak right outside the soft gap as seen in the DOS at $U=4$ and $W=30$ (Fig.~\ref{fig:DOS}). Since the mean distance between localized states in an energy window of the width $\Delta E$ centered at the energy $E$ is proportional to $(A(E)\Delta E)^{-1/d}$, the mean distance between localized states is shortest at the peak position. Thus, with decreasing $W$, the localized states at the peak position become extended before those at other energies, leading to the appearance of the low-lying extended excited states in the vicinity of the metal-insulator transition. Therefore the reentrant localization and the low-lying extended excited states may be ubiquitous in insulators with a soft gap.

\subsection{One dimension}\label{sec:1D}
In order to clarify how the unconventional scaling observed in 3D depends on the spatial dimensionality and is modified in lower dimensions, we further numerically analyze the Anderson-Hubbard model in 1D with the box distribution of $P_V$ in detail here.
\subsubsection{Hartree-Fock approximation}
\begin{figure}[h]
  \centering
  \includegraphics[width=0.425\textwidth,clip]{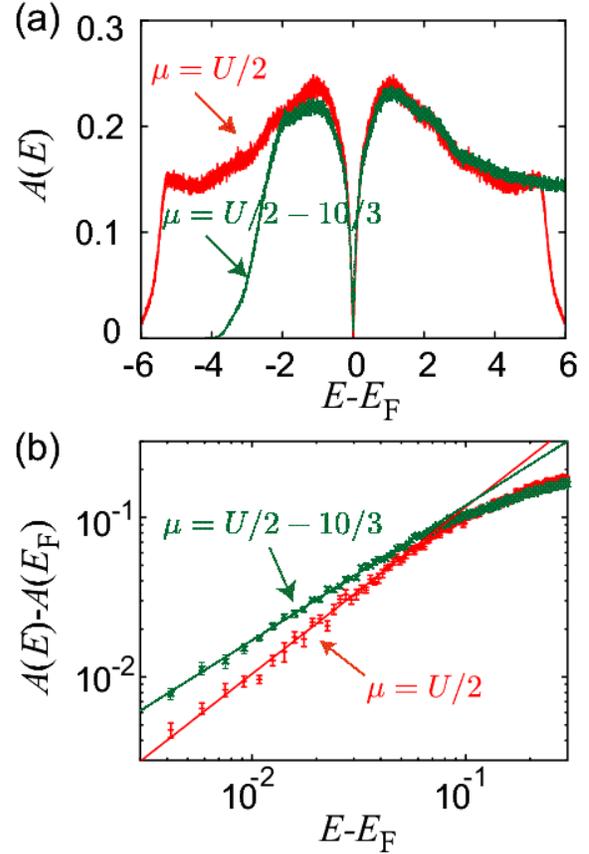}
  \caption{\coloronline DOS by HF in 1D at $t=1$, $U=10/3$, $W=20/3$ ($\Ns=14$) in the linear (a) and the logarithmic (b) plots. Electron filling is half-filled and hole-doped for $\mu=U/2$ and $\mu=U/2-10/3$, respectively. The straight lines in (b) are the scaling plots.}
  \label{fig:dos-1D-chem}
\end{figure}

\begin{figure}[h]
  \centering
  \includegraphics[width=0.35\textwidth,clip]{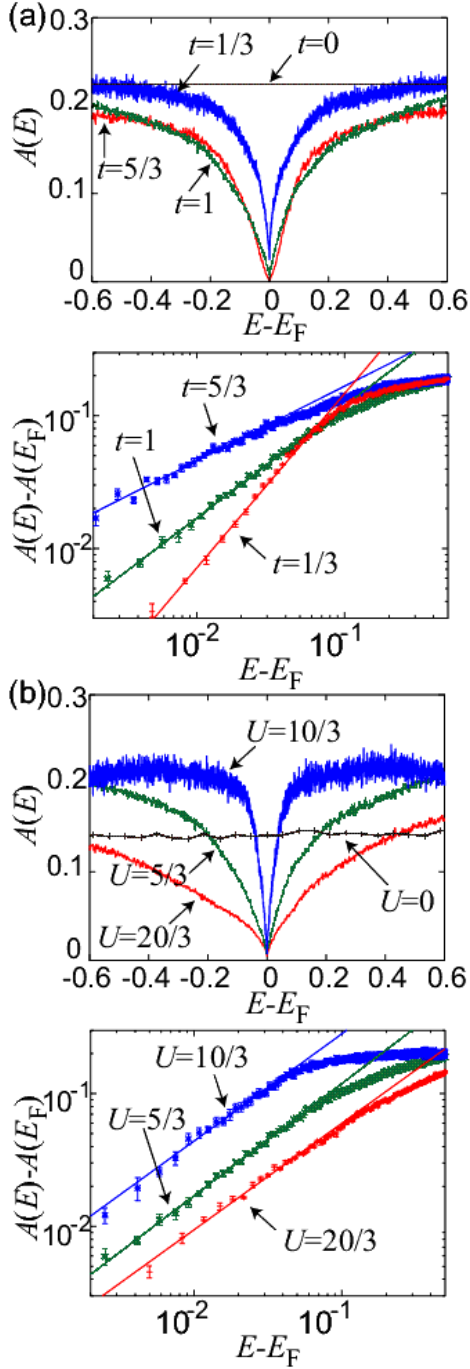}
	\caption{\coloronline Dependence of DOS on $t$ (panel (a)), $U$ (panel (b)) in 1D within the HF approximation ($\Ns=14$). The straight lines are the fit by power laws: (a) $U=10/3$, $W=20/3$, (b) $t=1$, $W=20/3$.}
  \label{fig:dos-1D-tU}
\end{figure}

We find the unconventional soft gaps also in 1D regardless of electron filling within the HF approximation. Figure~\ref{fig:dos-1D-chem} (a) shows the DOS for the hole-doped case as well as for the half-filled case. We employ the periodic boundary condition. Here, holes are doped with the chemical potential $\mu$ shifted by $-10/3$ from the half filling to increase the average distance between electrons to capture long-range asymptotic behavior easily. Surprisingly in contrast to the 3D case, they fit well with a power law $A(E)\propto |E-E_\mathrm{F}|^{\alpha}$ even at low energies as shown in Fig.~\ref{fig:dos-1D-chem} (b) regardless of electron filling. Although the Efros-Shklovskii theory predicts $A(E_\mathrm{F})>0$ in 1D (see \Eq~(\ref{eq:ES-scaling})), a logarithmic law of the soft gap was derived in 1D by considering excitations with multi electron-hole pairs as~\cite{Raikh87}
\begin{eqnarray}
	A(E) \propto \frac{1}{\log(E_0/|E-E_0|)}.\label{eq:ES-1D}
\end{eqnarray}
However, this conventional theory cannot explain the observed power law, which clearly indicates the existence of the unconventional mechanism of the soft gap. Although the gaps again shrink with the decreasing energy scale when $t$ or $U$ becomes smaller, the power law still holds when $t$ and $U$ are nonzero as shown in Fig.~\ref{fig:dos-1D-tU}.

\subsubsection{Exact diagonalization}\label{section:ED}
\begin{figure}
 \centering
 \includegraphics[width=0.425\textwidth,clip]{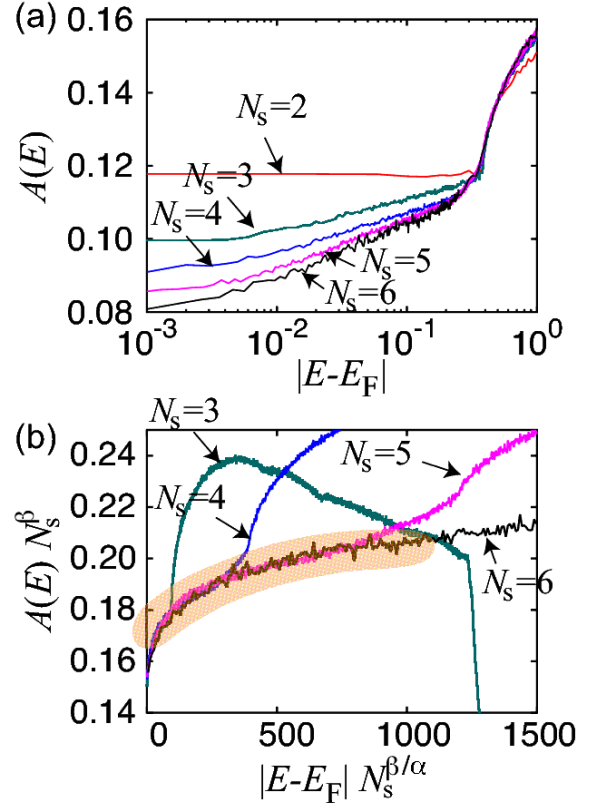}
 \caption{\coloronline (a) DOS in 1D with the exact diagonalization (open boundary condition): $t=1$, $U=10$, $W=10$, $\mu=U/2$, ($\Ns=2,3,4,5,6$). We average the DOS over $3.2\times10^7$ realizations of disorder for $\Ns=6$. (b) Scaling plot by $A(\epsilon, \Ns^{-1}) = \Ns^{-\beta} f(\epsilon \Ns^{\beta/\alpha})$ ($\alpha=0.075, \beta=0.375$). Shaded part shows the converged scaling curve.}
 \label{fig:ED-SCALING} 
 \label{fig:dos-ED-log}
\end{figure}

In order to investigate effects of quantum fluctuations not taken into account in the HF approximation, we further analyze the Anderson-Hubbard model by using the exact diagonalization in 1D. In Fig.~\ref{fig:dos-ED-log}, we further show the calculated DOS at $t=1$, $U=10$ and $W=10$. Because we need all the eigenstates to calculate the excitation spectra of each ensemble with a fixed configuration of randomness and the ensemble average over an extensive number of random configurations is required, we restrict the system size $\Ns$ to $\Ns\le6$. We find a dip in the DOS at $E_\mathrm{F}$, where the DOS decreases as the system size $\Ns$ increases. This indicates the formation of a soft Hubbard gap beyond the mean-field level. In order to clarify the scaling of the DOS, we employ a finite-size scaling analysis. We assume a scaling function, $A(\epsilon, \Ns^{-1}) = \Ns^{-\beta} f( \epsilon \Ns^{\beta/\alpha})=\epsilon^\alpha g( \Ns^{-\beta/\alpha} \epsilon^{-1})$ corresponding to $A(\epsilon, \Ns^{-1}=0) \propto  \epsilon^\alpha$ and $A(\epsilon=0, \Ns^{-1}) \propto \Ns^{-\beta}$ ($\epsilon = | E-E_\mathrm{F}|$). As shown in Fig.~\ref{fig:ED-SCALING}, the DOS well converges to this scaling function with $\alpha=0.075$ and $\beta=0.375$. Although a possible logarithmic scaling cannot be excluded because of the limitation in the system sizes of the present exact-diagonalization calculation, the exact-diagonalization results are consistent with the HF results and support a mechanism of the soft gap working beyond the mean-field level.

\begin{figure}[h]
  \begin{center}
  \includegraphics[width=0.425\textwidth,clip]{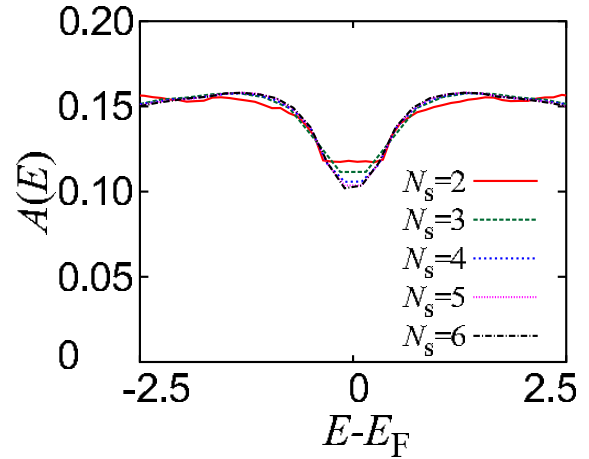}
  \end{center}
	\caption{\coloronline Linear plot of the same data as in Fig.~\ref{fig:dos-ED-log} with a lower energy resolution of $0.2t$.}
  \label{fig:ED-LOW-RESOLUTION}
\end{figure}

On the other hand, S. Chiesa \textit{et al.} found only a dip of the DOS in 2D with exact diagonalization~\cite{Chiesa08}. Namely, they found that the DOS is almost system-size independent within the energy resolution of the order of $0.1t$, which appears to disagree with the present result. However, the pseudogap observed in their study may well correspond to high-energy part of a soft gap in our analyses. Indeed, the soft gap is restricted to very low energies \textit{e.g.}, $|E-E_\mathrm{F}|<0.3t$ in our 1D study. At higher energies, the DOS is almost system-size independent and looks like a pseudogap if the energy resolution becomes poor (\textit{e.g.}, the energy resolution $\Delta E=0.2t$) as shown in Fig.~\ref{fig:ED-LOW-RESOLUTION} being consistent with their observation. Thus we believe that further analyses at lower energies with higher energy resolution will reveal soft gaps also in 2D. 

%% file: theory.tex
\section{Theory of Soft Hubbard Gap}\label{SEC:THEORY}
Our numerical observation of an unconventional soft gap urges us to explore an unconventional mechanism for its origin and an unprecedented low-energy excitations when all of the interaction, randomness and quantum effects are combined. In {\S}~\ref{sec:scaling}, we propose a multivalley energy landscape as the origin of the soft Hubbard gap and also propose the resultant scaling theory. In addition, we compare our theory with the numerical results in detail to test the mechanism. Although the scaling theory has been reported briefly in the previous letter,~\cite{Shinaoka08} we discuss in more detail for the benefit of its further extension: Although a continuous distribution of random potentials is assumed for the moment for the simplicity of discussion, we extend our scaling theory to discrete distributions of random potentials in {\S} \ref{sec:discrete-distribution}. Section~\ref{sec:long-range} is devoted to consideration of effects of the long-range part of the Coulomb interaction and extension of the scaling to the long-range Coulomb interaction. Finally, in {\S}~\ref{sec:scaling-DOS}, we summarize the scaling laws of the DOS obtained in this section.
\subsection{Origin and scaling theory}\label{sec:scaling}
In this section, we propose the origin of the soft Hubbard gap and construct its scaling theory. For simplicity without loss of generality, we restrict ourselves to a single-particle excitation for the electron side, namely, $E>E_\mathrm{F}$. We consider the case of $r_\mathrm{int} \ll \xi$, where $r_\mathrm{int}$ is the range of the interaction in the model; $r_\mathrm{int}=0$ for the Anderson-Hubbard model. Even for $r_\mathrm{int}=0$, when $t$ becomes nonzero, virtual hopping of electrons generates intersite effective interaction $U_{ij}$ though the zero point fluctuation, which exponentially decreases with the mutual distance $|r_i-r_j|$ as
\begin{eqnarray}
	U_{i j}\propto\exp(-b |r_i-r_j|), \label{eq:U-r}
\end{eqnarray}
where $b$ is proportional to the inverse of the localization length. This effect is not considered in the ES theory because it regards electrons as classical particles. Within the HF approximation, the interaction energy (the second term of the Anderson-Hubbard Hamiltonian, \Eq~(\ref{eq:ham})) is decoupled as
\begin{eqnarray}
\Bigl\langle U \sum_{i} n_{i \uparrow}n_{i \downarrow} \Bigr\rangle\!\!&=&\!\!\sum_{n,m} \Bigl\{U \sum_{i} (|\phi_m(r_i;\uparrow)|^2|\phi_n(r_i;\downarrow)|^2\nonumber \\
&&+|\phi_m(r_i;\downarrow)|^2|\phi_n(r_i;\uparrow)|^2)\Bigr\},
\end{eqnarray}
where the orbital indices $n$ and $m$ run over all the occupied single-particle wave functions. Thus, from \Eq~(\ref{eq:localization-length}), $b$ is given by $b=2/\xi$ within the HF approximation.

By assuming the self-averaging of the DOS, the DOS averaged over the sites is obtained as
\begin{eqnarray}
 A(E) &=&  {\Bigl\langle{\int}_{-\infty}^{\infty} P_V(V_1)A_1(E,V_1) {\rm d} V_1 \Bigr\rangle}_{\{ V_{\overline{1}} \}}, 
\end{eqnarray}
where the symbol $\{ V_{\overline{1}}\}$ denotes a set of random potentials $V_i$ except for $V_1$. Note that $A_1 (E, V_1)$ is the local DOS projected on the site $1$ under the condition of the fixed $V_1$ at the site $1$. This local DOS $A_1(E, V_1)$ implicitly depends on ${\{ V_{\overline{1}}\}}$. Here we decompose the average over the random potential into the two steps; namely, the average over $V_1$ as described by $\int P_V(V_1) d V_1$ at a fixed configuration of $\{ V_{\overline{1}}\}$ and the subsequent average with respect to $\{V_{\overline{1}}\}$. In the following, we first examine $V_1$-dependence of the single-particle excitation energy, namely $A_1 (E, V_1)$ and obtain the $V_1$-averaged local DOS. Then we average the $V_1$-averaged local DOS over $\{V_{\overline{1}}\}$ to obtain the scaling of the soft Hubbard gap.

\begin{figure}
 \centering
 \includegraphics[width=0.375\textwidth,clip]{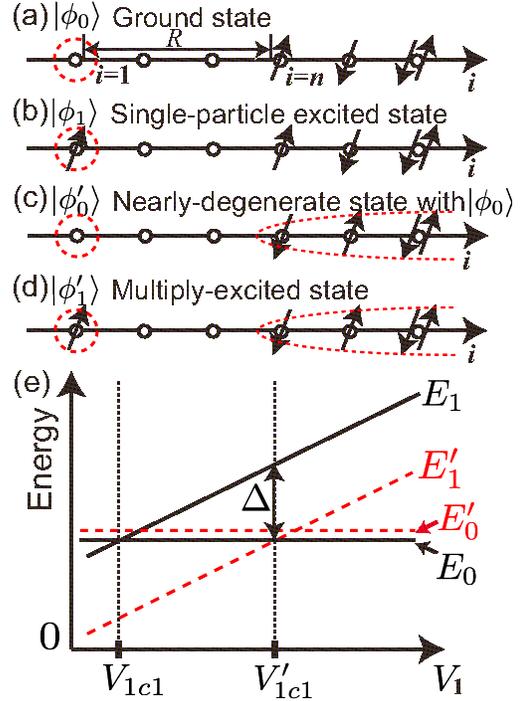}
 \caption{\coloronline Schematic illustration of (a) the ground state, (b) a single-particle excited state, (c) a nearly-degenerate state with the ground state and (d) a multiply-excited state. (e) Schematic of $V_1$ dependence of excitation energies.}
 \label{fig:level-cross}
\end{figure}
First, we discuss $V_1$-dependence of $A_1 (E, V_1)$ for fixed ${\{ V_{\overline{1}}\}}$. When $V_1$ decreases, the ground-state occupation of the site 1 changes from 0 to 1 and then from 1 to 2 at $V_{1c1}$ and $V_{1c2}$, respectively. A possible and typical ground state $|\phi_0\rangle$ at $V_1>V_{1c1}$ is illustrated in Fig.~\ref{fig:level-cross}(a), where the site 1 is empty. Here the total particle number is $\Ne=N_a$ and the energy $E_{0}(V_1)$. Near $V_{1c1}$ but for $V>V_{1c1}$, a single-particle excited state $|\phi_1\rangle$ with $\Ne=N_a+1$ and the energy $E_{1}(V_1)$ is defined by the electron configuration fixed to be the same as $|\phi_0\rangle$ except for an addition of an electron at the site 1, as is illustrated in Fig.~\ref{fig:level-cross}(b). In the interacting case, because $c_{i\sigma}^\dagger|\phi_0\rangle$ is not necessarily an eigenstate, there may be several eigenstates that have nonzero matrix elements with $c_{i\sigma}^\dagger | \phi_0 \rangle$, namely, single-particle excited states. However, for simplicity, we discuss the lowest single-particle excited state, $|\phi_1\rangle$, which dominates at low energies. One might think that $|\phi_1\rangle$ becomes the ground state below $V_{1c1}$, where $\Ne=N_a+1$. In this case, however, the single-particle excitation gap $E_1-E_0$ vanishes at $V_{1c1}$ leading to the absence of the gap in the $V_1$-averaged DOS. Indeed, this is what happens in the Anderson insulator at $U=0$, where the DOS exhibits no soft gap. Thus the numerical evidences of the soft gaps indicate that $|\phi_1\rangle$ as a single-particle excited state is excluded by the electron correlation. 

In the ES theory, single-particle excited states are excluded from low energies by the ground-state stability condition against an electron-hole excitation. This mechanism, however, cannot be attributed to the formation of the soft Hubbard gap, because this excitonic effect is negligible in the case of the short-range interaction. We now consider the ground-state stability against excitations of more complex form. We assume a multivalley energy landscape, which may be characteristic to random systems. Namely, we assume that there exist many arbitrarily-low-energy excited states whose occupations are the same with those of $|\phi_0\rangle$ at the site $1$ but whose configurations are globally different on other sites. In Fig.~\ref{fig:level-cross}(c), we illustrate a state $|\phi_0^\prime\rangle$ at a local minimum $E_0^\prime$ nearly degenerate with $|\phi_0\rangle$. Here the configurations of $|\phi_0^\prime \rangle$ are relaxed from $|\phi_0\rangle$ not only at the occupied site $n$ nearest to the site $1$ at the distance $R$ but also at farther sites ($>R$). Figure~\ref{fig:level-cross}(d) shows a single-particle excited state $|\phi_1^\prime \rangle$ from $|\phi_0^\prime\rangle$ with the energy $E_{1}^\prime$ and the site-$1$ occupancy identical with $|\phi_1 \rangle$. Here, the two nearly-degenerate states, $|\phi_1\rangle$ and $|\phi_1^\prime\rangle$ are separated by a barrier, where multi-particle relaxation is required to reach from one to the other. Now $E_1$ is given by $(V_1-E_\mathrm{F}) + \sum_i U_{1 i}+E_0$, where $U_{1 i}$ is the interaction energy between electrons on the site $1$ and those on the site $i$. Note that the interaction energy $|U_{1i}|$ is nonzero only when the site $i$ satisfies $R \le |i-1| \lesssim R+\xi$ because of the localization (Keep in mind that $R$ is the distance to the nearest neighbor electrons from the site $1$). On the other hand, $E_1^\prime$ is given by $E_1^\prime=(V_1-E_\mathrm{F}) + \sum_i U^\prime_{1 i}+E_0^\prime$, where $U_{1 i}^\prime$ is again the interaction energy between electrons on the site $1$ and those on the site $i$ in the state $|\phi_0^\prime\rangle$. Because $E_0 \simeq E_0^\prime$ and the configurations of $|\phi_1^\prime\rangle$ on these sites are different from those of $|\phi_1\rangle$, $E_1^\prime$ is different from $E_1$ typically by the amount as much as $|U_{1n}|$. Now, out of many possible $|\phi_0^\prime \rangle$s, one can choose $|\phi_0^\prime\rangle$ so that the energy $E_1^\prime$ is lower than $E_1$ by the amount $|U_{1 n}|$. Then $|\phi_1^\prime\rangle$ is indeed a state that is obtained by a multiple excitation from $|\phi_0\rangle$. This means that $\langle \phi_1^\prime | c^\dagger_{i\sigma} | \phi_0 \rangle=0$ and $|\phi_1^\prime\rangle$ is orthogonal to the single-particle excitations.

Now we assume linear dependence of the excitation energies, $E_1(V_1)$ and $E_1^\prime(V_1)$ as functions of $V_1$. Then $E_1^\prime(V_1)$ and $E_0(V_1)$ crosses at $V_1=V_{1c1}^\prime$ and for $V_1<V_{1c1}^\prime$ the ground state becomes $|\phi^\prime_1\rangle$ as illustrated in Fig.~\ref{fig:level-cross} (e). Note that the excitation energy $E_1^\prime-E_1$ is negative in the vicinity of $V_1=V_{1c1}^\prime$. For $V_1>V_{1c1}^\prime$, the state $|\phi^\prime_1\rangle$ is not counted in the DOS, because this state is not a single-particle excitation of $|\phi_0\rangle$, but rather a multiply-excited state. As a result, the single-particle excitation energy $E_1-E_0$ has the lower bound at $V_1=V_{1c1}^\prime$. In other words, the energy difference $\Delta = |E_1(V_{1c1}^\prime) - E_1^\prime(V_{1c1}^\prime) |$ should be the lowest energy of single-particle excitations counted in $A$ near $V_1=V_{1c1}^\prime$ in the region $V_1>V_{1c1}$. Therefore the $V_1$-averaged local DOS has a gap of $\Delta$ in contrast to the non-interacting case as
\begin{equation}
 {\small \int}_{V_{1c1}^\prime}^{\infty} P_V(V_1)A_1(E,V_1) {\rm d} V_1 \propto H_\mathrm{s} (E-E_\mathrm{F}-\Delta), \label{eq:local_A}
\end{equation}
where $H_\mathrm{s}$ is the Heaviside step function. The same argument applies around $V_1=V_{1c2}$.

Here we make an additional comment. One might think that, as in the ES theory, it could be possible to lower the energy of $|\phi_1 \rangle$ from $E_1$ to $E_1^\prime$ by relaxing \textit{local} electronic configurations only near the site $n$. It, however, always increases the energy of the electrons on the sites other than the site $1$, because they have already been optimized in the ground state. Thus a global reconstruction is required to lower the energy.

Next, in order to derive the scaling of the density of states, we discuss the distribution function of $\Delta$ with respect to ${\{ V_{\overline{1}}\}}$. From the above discussion, it is reasonable to assume that $\Delta$ depends on ${\{ V_{\overline{1}}\}}$ only through $R$. Under this assumption, from \Eq~(\ref{eq:U-r}), $\Delta$ scales as
\begin{eqnarray}
 \Delta(R) &=&  a\exp(-b R), \label{eq:D-R}
\end{eqnarray}
where $a$ and $b$ ($\propto \xi^{-1}$) are non-universal positive constants. It is clear that $a$ is proportional to $U$ when $U$ is small, and $b$ diverges for $t \rightarrow 0$. Hereafter we neglect logarithmic corrections. The localization lengths may fluctuate between the site $1$ and the site $n$. However, this fluctuation of the localization lengths does not affect the asymptotic behavior of $\Delta(R)$, \Eq~(\ref{eq:D-R}), because of the self averaging of the localization length between the site $1$ and the site $n$ in the limit of $R\rightarrow +\infty$.

\begin{figure}
 \centering
 \includegraphics[width=0.425\textwidth,clip]{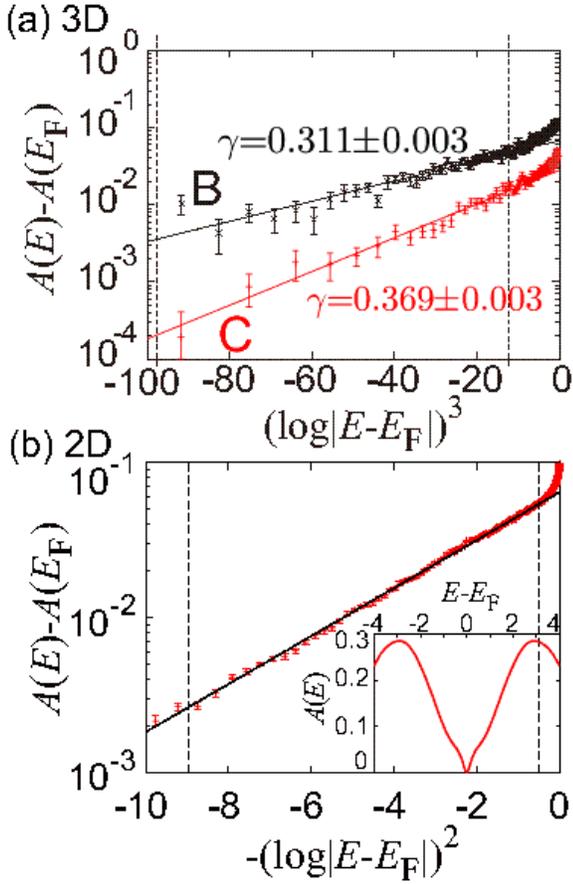}
 \caption{\coloronline (a) Scaling plot of DOS in 3D at parameters B and C depicted in Fig.~\ref{fig:pd}: B ($t=1,~U=4,~W=30$), C ($t=1,~U=6,~W=5$). We employ Lorentz broadening with a broadening factor $1.25\times10^{-3}$ and $6.25\times10^{-4}$ for A and B, respectively. The broken lines denote $|E-E_\mathrm{F}|=10^{-2}$ and $10^{-1}$. The DOS fits well with $A(E) \propto\exp(-(-\gamma\log| E-E_\mathrm{F}|)^3 )$ shown by the fitting lines for $10^{-2}<|E-E_\mathrm{F}|<10^{-1}$. (b) Scaling plot of DOS in 2D: $t=1, U=6, W=5$ with the Gaussian distribution of $P_V$. The inset is a linear plot of the same data. The black solid curves are fit by the predicted scaling,  $A(E) \propto\exp(-(\gamma\log| E-E_\mathrm{F}|)^2)$. If $A(E)$ followed $A(E) \propto\exp(-(\gamma\log| E-E_\mathrm{F}|)^2)$, the data in the panel (b) would follow a straight line. The broken lines denotes $|E-E_\mathrm{F}|=0.05, 0.5$.}
 \label{fig:dos-scaling-HF-3D}
\end{figure}

On the other hand, the probability distribution of $R$ follows 
\begin{eqnarray}
 P(R) &=&  a^\prime R^{d-1} \exp(-b^\prime R^d), \label{eq:P(R)}
\end{eqnarray}
at long distances, where $a^\prime$ and $b^\prime$ are non-universal positive constants again.  Equation~(\ref{eq:P(R)}) means that the probability of formation of a large void of electrons around the site $1$ is exponentially rare. Equations~(\ref{eq:D-R}) and (\ref{eq:P(R)}) lead to
\begin{eqnarray}
 Q(\Delta)&=&P(R(\Delta)) \left| \frac{dR}{d\Delta} \right| \nonumber \\
 &\propto& (-\log \Delta)^{d-1} \Delta^{-1} \exp\left({-\frac{b^\prime}{b^d} (-\log \Delta)^d}\right) \nonumber \\
 &\simeq& \Delta^{-1} \exp\left({-\frac{b^\prime}{b^d} (-\log \Delta)^d}\right), \label{eq:P(D)}
\end{eqnarray}
where $Q(\Delta)$ are the distribution function of $\Delta$. Here we neglect the logarithmic correction term $(-\log \Delta)^{d-1}$. Because the DOS at the energy $E$ is proportional to the probability of $\Delta \le |E-E_\mathrm{F}|$, we obtain the scaling of the DOS from Eqs~(\ref{eq:local_A}) and (\ref{eq:P(D)}) as
\begin{eqnarray}
 A(E)&\propto&{\small \int}^{|E-E_\mathrm{F}|}_{0} {\rm d} \Delta Q (\Delta) \nonumber \\
 &\propto& \exp\left({-\frac{b^\prime}{b^d} (-\log |E-E_\mathrm{F}|)^d}\right). \label{eq:DOS-HF}
\end{eqnarray}
For $d=1$, because the exponential and the logarithm functions cancel each other, this scaling reduces to a power law with a non-universal exponent, which is consistent with the observed power-law scaling in 1D: $A(E) \propto {|E-E_\mathrm{F}|}^{b^\prime/b}$. Non-universal power-law distributions of energies are common in Griffith phases~\cite{Griffith69}. Equation~(\ref{eq:D-R}) indicates that $\Delta$, namely the gap vanishes as $t$ or $U$ vanishes because of $a\rightarrow 0$ or $b\rightarrow \infty$. Furthermore, the exponent $\alpha=b^\prime/b$ is expected to decrease as $t$ becomes smaller because of the reduction of $\xi$. These predictions are consistent with our HF results in 1D as shown in Fig.~\ref{fig:dos-1D-tU} (a). It should be mentioned that the power law for 1D in the present theory is in sharp contrast with the ES theory, because the ES theory predicts the logarithmic scaling of the DOS as \Eq~(\ref{eq:ES-1D}). For $d>1$, our scaling predicts that the DOS vanishes faster than any power law, being consistent again with our HF study in 3D. In Fig.~\ref{fig:dos-scaling-HF-3D} (b), we show a scaling plot of the DOS in 3D by our scaling given by \Eq~(\ref{eq:DOS-HF}). Indeed, the DOS fits well with our scaling in 3D at low energies \textit{e.g.,} $|E-E_\mathrm{F}|<0.1$, where the DOS does not follow a power law.

Because our scaling has explicit dimension dependence, we further test our scaling in 2D within the HF approximation with the Gaussian distribution of $P_V$. Figure~\ref{fig:dos-scaling-HF-3D} (b) shows the DOS with the system size $10\times10$ and the periodic-boundary condition at $t=1$, $U=6$, $W=5$. We averaged the DOS over as many as $4.8\times10^4$ realizations of random potentials to obtain the high-resolution data. As shown in Fig.~\ref{fig:dos-scaling-HF-3D} (b) the DOS fits well with our scaling obtained from \Eq~(\ref{eq:DOS-HF}) rather than a power law for $|E-E_\mathrm{F}|<0.5$.

\begin{figure}[bth]
  \begin{center}
  \includegraphics[width=0.425\textwidth,clip]{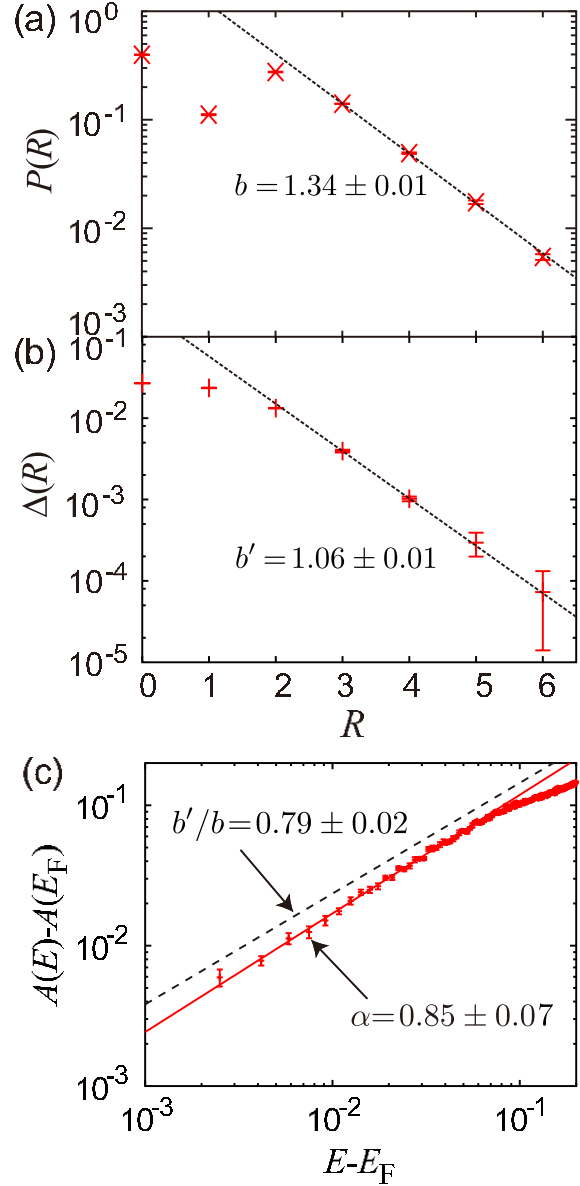}
  \end{center}
  \caption{\coloronline Numerical estimates of $P(R)$ and $\Delta (R)$ at $t=1$, $U=10/3$, $W=20/3$ and $\mu=U/2-10/3$. Fittings by \Eqs~(\ref{eq:D-R}) and (\ref{eq:P(R)}) as are shown in (a) and (b) by dotted lines give $b^\prime = 1.06 \pm 0.01$ and $b=1.34 \pm 0.01$. The ratio $b^\prime/b$ (dashed lines in (c)) well agrees with $\alpha$ obtained by the fitting (solid line) of the DOS in (c).}
  \label{fig:dos-1D-param}
\end{figure}

In addition to the qualitative consistency between our scaling and the numerical results, in Fig.~\ref{fig:dos-1D-param}, we show a further numerical evidence of our theory on the quantitative level for 1D. Figure \ref{fig:dos-1D-param} (a) and (b) shows $\Delta(R)$ and $P(R)$ calculated by the following procedure: First, we obtain the ground state for each realization of random potentials. We construct the lowest single-particle excited state by adding one electron to the lowest unoccupied orbital. Next we optimize the mean fields by the iterative scheme starting from those of the single-particle excited state with $\Ne$ fixed. Then $\Delta$ is obtained as the difference of these two excitation energies. We calculate $R$ as the distance between the center of the lowest unoccupied orbital, $r$ and those of the occupied orbitals nearest to $r$ in the ground state. We define the center of the orbital as the site that has the maximum weight. Fitting by \Eqs~(\ref{eq:D-R}) and (\ref{eq:P(R)}) gives $b^\prime = 1.06 \pm 0.01$ and $b=1.34 \pm 0.01$. Estimated exponent of $b^\prime/b=0.79\pm0.02$ is in good agreement with $\alpha=0.85 \pm 0.07$ obtained directly from the DOS as shown in Fig.~\ref{fig:dos-1D-param} (c). This is a numerical evidence for the quantitative validity of our theory. 

In the above discussion, under the assumption of a multivalley energy landscape, we have successfully constructed the scaling theory which is consistent with all the numerical results. In the presence of a multivalley energy landscape, there are many excited states whose configurations are globally different from those of the ground state, which is a non-trivial combined effect of the electron correlation and randomness. Indeed, in the field of spin glasses, it is known that a multivalley energy landscape emerges concurrently with replica symmetry breaking (RSB) within a spin-glass phase. Extensive studies on replica symmetry breaking (RSB) in finite dimensions have been carried out especially in classical Ising models (Ising Edward-Anderson model).~\cite{Sasaki07} On the other hand, our Hartree-Fock results indicate that a multivalley energy landscape exists over the whole insulating phases for $d=1$, $2$ and $3$. Furthermore, although the exact-diagonalization results are restricted to the strongly-localized region in 1D because of the severe limitation of system sizes, the exact-diagonalization results indicate robustness of the multivalley energy landscape against quantum fluctuations beyond the mean-field level. Further theoretical studies along this line are intriguing future subjects.

Our conclusion disagrees with the DMFT results~\cite{Dobrosavljevic97} as well as several other mean-field studies~\cite{Dobrosavljevic03, Byczuk05} which support absence of the soft gaps. This may be because they ignore spatial correlations. The latter ignores inhomogeneity of the electronic structures. Indeed, a DMFT study improved by partially taking account of the intersite self-energy retrieves the suppression of the DOS near $E_\mathrm{F}$ to some extent~\cite{Song08}.

In contrast to our scaling, the power law was proposed to interpret the photoemission experiments~\cite{Kim06, Maiti07}. However, as shown in Figs.~\ref{fig:DOS} and \ref{fig:dos-scaling-HF-3D}, the asymptotic behavior of the DOS is restricted to low energies, namely $|E-E_\mathrm{F}|<0.1t$ and high-energy part of the DOS seems as if it approximately follows a power law with a non-universal exponent in our 3D HF study. Because hopping integrals between $d$ orbitals on nearest-neighbor Ru atoms are on the order of $0.1$ eV for SrRuO$_3$~\cite{Mazin00}, the asymptotic behavior of the soft Hubbard gap may be restricted to the energy region lower than $10$ meV. Thus photoemission experiments with the energy resolution on the order of $1$ meV are desired to observe the present asymptotic behavior clearly. Although this level of resolution has become possible recently, such high resolution has not been utilized so far for the present purpose in the literature~\cite{Kim06, Maiti07}. We believe that our paper provides incentive for such high-resolution photoemission experiments. In addition to the high-resolution photoemission, other experiments accessible to low energies are highly desired for experimental confirmation of the present theory. For example, the DC transport measurement is suitable for investigating the density of states in insulating phases at low energies \textit{i.e.} $T<300~\mathrm{K}$ ($k_\mathrm{B} T \lesssim 30~\mathrm{meV}$). Actually, in {\S} \ref{sec:transport}, we discuss the temperature dependence of the DC resistivity in the presence of the soft Hubbard gap.

\subsection{Extension of scaling theory to discrete distributions of random potentials}\label{sec:discrete-distribution}
Although we assume a continuous distribution $P_V$ in the previous sections, the obtained scaling, \Eq~(\ref{eq:DOS-HF}) does not depend on the distribution. Even for a discrete distribution of a binary alloy form, one obtains the same scaling as that for continuous distributions by modifying the above discussion slightly. In the above discussion, we first average the single-particle excitation spectrum over $V_1$ at fixed configurations of ${\{ V_{\overline{1}}\}}$,  which is a set of random potentials $V_i$ except for $V_1$. For a continuous distribution, as described in \Eq~(\ref{eq:local_A}), the single-particle excitation energy distributes continuously above a lower bound $\Delta (R)$ depending on ${\{ V_{\overline{1}}\}}$ through $R$. Here $R$ is the minimum distance from the site $1$ to the occupied sites in the ground state. This continuous distribution is a key point for obtaining our scaling. For a discrete distribution, however, the single-particle excitation energy distributes discretely above $\Delta (R)$, because $V_1$ is discrete. Thus we need an extension of our consideration here. We further average the single-particle excitation spectrum over ${\{ V_{\overline{1}}\}}$ with a certain common value of $R$. Then the distribution of the single-particle excitation energy becomes continuous above $\Delta (R)$, because of fluctuations of the electronic structures at the distance $R$ from the site $1$ in the ground states. After a further average over $R$, one obtains the same scaling as that for the continuous distribution.

\begin{figure}[bth]
  \begin{center}
  \includegraphics[width=0.425\textwidth,clip]{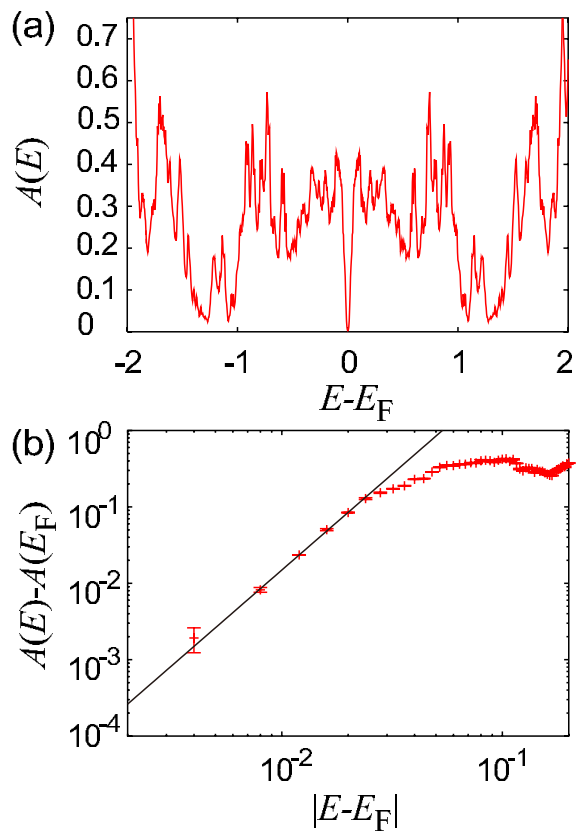}
  \end{center}
	\caption{\coloronline Density of states in 1D within the Hartree-Fock approximation with 24 sites: $t=1, U=2, W=2$. (a) Linear plot. (b) Double logarithmic plot. The solid line in (b) is the fit with a power law.}
  \label{fig:1D-bimodal}
\end{figure}

We show a numerical evidence in 1D within the Hartree-Fock approximation in Fig.~\ref{fig:1D-bimodal}. We employ the bimodal distribution: $V_i=\pm W$ with the symmetric probability $P(W)=1/2$ and $P(-W)=1/2$. Although the density of states has a complex peak structure due to the discreteness of the bimodal distribution, the density of states clearly exhibits a soft gap near the Fermi energy as shown in Fig.~\ref{fig:1D-bimodal} (a). Indeed, as shown in Fig.~\ref{fig:1D-bimodal} (b), the soft gap follows a power law, which is consistent with the above discussion.

\subsection{Effects of the long-range part of the Coulomb interaction}\label{sec:long-range}
Even near metal-insulator transitions where the screening is strong, the Coulomb interaction remains long ranged in insulating phases, though its amplitude may be small. Thus the scaling of the DOS deviates from \Eq~(\ref{eq:DOS-HF}) in real materials, as the long-range part becomes dominant at low energies. Note that \Eq~(\ref{eq:DOS-HF}) is the asymptotic scaling for the model with short-range interaction only. In this section, we discuss effects of the remaining long-range part of the Coulomb interaction.

Owing to the long-range part of the Coulomb interaction, Efros \textit{et al.} showed that a ground-state stability against an exciton excludes low-energy single-particle excitations and generates power-law soft gaps as $A(E) \propto |E-E_\mathrm{F}|^{d-1}$ (see \Eq~(\ref{eq:ES-scaling})).~\cite{Efros-Shklovskii} However, since they considered the stability condition only against electron-hole excitations, the DOS may vanish faster than the ES scaling at lower energies because of stability conditions against further many-particle excitations.  Indeed, Efros obtained a correction of the scaling in 3D by considering multi-electron-hole excitations as~\cite{Efros76}
\begin{eqnarray}
	A(E) &\propto& \exp(-(\epsilon_0/\epsilon)^{1/2}), \label{eq:corr-Efros}
\end{eqnarray}
where $\epsilon=|E-E_\mathrm{F}|$ and $\epsilon_0$ is a constant. For $d=1,2$, he claimed that there is no further correction in ES scaling.

However, since our scaling even for the short-range interaction shows faster decay of the DOS than any power law in the presence of a multivalley structure, it is natural to infer that the DOS should decay even faster in the presence of the long-range interaction and certainly beyond the ES theory at low energies instead of \Eq~(\ref{eq:corr-Efros}). Thus we extend our scaling theory to the case of the long-range Coulomb interaction. For the long-range Coulomb interaction, \Eqs~(\ref{eq:D-R}) and (\ref{eq:P(R)}) are modified as
\begin{eqnarray}
 \Delta(R) &=&  aR^{-1}, \label{eq:D-R-L}\\
 P(R) &=&  a^\prime R^{d-1}\exp(-b^\prime R^d). \label{eq:P(R)-L}
\end{eqnarray}
From \Eqs~(\ref{eq:D-R-L}) and (\ref{eq:P(R)-L}), we obtain
\begin{eqnarray}
 Q(\Delta) &=& P(R) \left| \frac{d R}{d \Delta} \right| \nonumber\\
 &=& a^d a^\prime \Delta^{-d-1} \exp(-a^d b^\prime \Delta^{-d}).
\end{eqnarray}
In a manner similar to the short-range interaction (\Eq~(\ref{eq:DOS-HF})), the scaling of the DOS is given by
\begin{eqnarray}
	A(E) &\propto& \int^{|\epsilon|}_0 Q(\Delta) d \Delta \nonumber\\
  &\propto& \int^{|\epsilon|}_0  \Delta^{-d-1} \exp(-a^d b^\prime \Delta^{-d}) d\Delta \nonumber\\
  &=& \frac{1}{a^d b^\prime d} \exp(-a^d b^\prime \epsilon^{-d}) \nonumber\\
  &\propto& \exp(-a^d b^\prime \epsilon^{-d}).\label{eq:DOS-HF2}
\end{eqnarray}
Indeed, this scaling shows faster decay of the DOS than \Eq~(\ref{eq:corr-Efros}) proposed by Efros for 3D. Moreover, \Eq~(\ref{eq:DOS-HF2}) reveals that the ES scaling must be modified even for $d=1,2$.

\subsection{Summary of scaling laws}\label{sec:scaling-DOS}
In Table~\ref{table:scaling-DOS}, we summarize the scaling laws of the DOS in four kinds of models: (A)/(B) short-range interaction without/with a multivalley energy landscape, (C)/(D) long-range interaction without/with a multivalley energy landscape. Now we consider how the scaling of the soft gap depends on the energy for the case (D), which corresponds to realistic materials. At energies higher than the energy scale of the long-range part, the formation process of the soft gap is dominated by the short-range part. Thus, in this energy region, the soft gap follows the scaling of the soft Hubbard gap (\Eq~(\ref{eq:DOS-HF})). As the long-range part of the Coulomb interaction becomes dominating at low energies, the scaling of the DOS crosses over from \Eq~(\ref{eq:DOS-HF}) to the ES scaling \Eq~(\ref{eq:ES-scaling}) and further to \Eq~(\ref{eq:DOS-HF2}). It should be mentioned that the intermediate scaling laws may not always be observed clearly.

\begin{table*}[tb]
 \centering
 \begin{tabular}{l!{\bvline{1.5pt}}llc}\bhline{1.5pt}
  \multicolumn{1}{c!{\bvline{1.5pt}}}{Model} &   \multicolumn{2}{c}{Scaling of DOS} & Energy \\ \bhline{1.5pt}
  {\small (A) Short-range interaction } &$\simeq A_0>0$ & &\\ 
  {\small \textit{without} multivalley energy landscape}  & & &\\\hline
  {\small (B) Short-range interaction} &\cellcolor{lightgray} $\exp\left({-\gamma(-\log \epsilon)^d}\right)$ & \Eq~(\ref{eq:DOS-HF}) &\\ 
  {\small \textit{with} multivalley energy landscape}  &  & &\\ \bhline{1.5pt}
  {\small (C) Long-range Coulomb interaction}&$\epsilon^{d-1}$ &\Eq~(\ref{eq:ES-scaling}) & (HEs)\\
  {\small \textit{without} multivalley energy landscape}    &$\exp(-(\epsilon_0/\epsilon)^{1/2})~(\mbox{3D})$ & \Eq~(\ref{eq:corr-Efros}) & (LEs)\\ \hline
                          &\cellcolor{lightgray}(D.1) $\exp\left({-\gamma(-\log \epsilon)^d}\right)$ & \Eq~(\ref{eq:DOS-HF}) & (HEs) \\ \cdashline{2-4}[.4pt/1pt]
  {\small (D) Long-range Coulomb interaction}&(D.2) $\epsilon^{d-1}$ & \Eq~(\ref{eq:ES-scaling})  & $\downarrow$\\
  {\small \textit{with} multivalley energy landscape}    &   (D.3) $\exp(-(\epsilon_0/\epsilon)^{1/2})~(\mbox{3D})$ & \Eq~(\ref{eq:corr-Efros}) & $\downarrow$\\
                          &\cellcolor{lightgray}(D.4) $\exp(-\beta \epsilon^{-d})$ & \Eq~(\ref{eq:DOS-HF2}) & (LEs)\\ \bhline{1.5pt}
 \end{tabular}
 \caption{Summary of scaling laws of DOS four kinds of models: (A)/(B) short-range interaction without/with a multivalley energy landscape, (C)/(D) long-range interaction without/with a multivalley energy landscape. Shaded part denotes the novel scaling laws obtained in this paper. The scaling of the case (C.2) is expected to appear in energies lower than that of the case (C.1). The case (D.1) corresponds to the energy scale where the short-range part of the Coulomb interaction is dominant, while the cases (D.2-4) correspond to that where the long-range part is dominant. In the cases of the long-range interaction, scaling crossovers are expected as a function of the energy. Abbreviations are: High energies (HEs); Low energies (LEs).}
 \label{table:scaling-DOS}
\end{table*}

%% file: transport.tex
\section{Transport Properties}\label{sec:transport}
In this chapter, we discuss the temperature dependence of the DC resistivity in the presence of the soft Hubbard gap or the soft Coulomb gap. Although the transport coefficient is determined by the two-particle (electron-hole) correlations and is not necessarily identical with the single-particle excitations measured in the DOS, the DC transport measurement is a useful and good tool for investigating the DOS at low energies with high resolution in insulating phases if the transport is dominated by independent single-particle excitations.
\subsection{Temperature dependence of the DC resistivity}
\subsubsection{Insulators with a hard gap}
When a gap is still open for weak disorder, thermally-excited carriers dominate the conduction and the temperature dependence of the DC resistivity follows the Arrhenius law as
\begin{eqnarray}
 \rho &=& \rho_0 \exp\left(\frac{T_0}{T}\right),
\end{eqnarray}
where $T_0$ is a constant. 

\subsubsection{Anderson insulators without soft gaps}
If a gap is closed for stronger disorder, localized states near $E_\mathrm{F}$ dominates the conduction. Mott showed that at sufficiently low temperatures, conduction results from electron hopping between localized state within a narrow band near $E_\mathrm{F}$, which is called variable-range hopping (VRH)~\cite{mott66}. He also showed that provided there is a non-vanishing DOS at $E_\mathrm{F}$, the temperature dependence of the DC resistivity exhibits universal behavior as
\begin{eqnarray}
 \rho &=& \rho_0 \exp\left[\left(\frac{T_0}{T} \right)^{1/(d+1)}\right].
\end{eqnarray}

\subsubsection{Anderson insulators with soft gaps}
Because VRH explicitly depends on the DOS near $E_\mathrm{F}$, soft gaps modify the temperature dependence of the resistivity qualitatively. Now we discuss scaling of the DC resistivity in the presence of a soft gap. As discussed in \S~\ref{sec:long-range}, the scaling of the DOS crosses over from \Eq~(\ref{eq:DOS-HF}) to the ES scaling \Eq~(\ref{eq:ES-scaling}) and further to \Eq~(\ref{eq:DOS-HF2}) as the long-range part of the Coulomb interaction becomes dominating at low energies. With this evolution of the crossover, $\rho$ also crossovers {as a function of the temperature}. In the following, we discuss scaling of the DC resistivity for each energy region separately.

{\bf\small (1) Energy regions (D.2) and (D.3) in Table~\ref{table:scaling-summary}}

Efros \textit{et al.} ~\cite{Efros-Shklovskii} showed that in the presence of the soft Coulomb gap with the power law $A(E)\propto |E-E_\mathrm{F}|^{d-1}$ (\Eq~(\ref{eq:ES-scaling})), the temperature dependence of the DC resistivity is modified regardless of the spatial dimension $d$ as
\begin{eqnarray}
 \rho &=& \rho_0 \exp\left[\left(\frac{T_0}{T} \right)^{1/2}\right]. \label{eq:rho-ES}
\end{eqnarray}
Even in the case (D.3) of Table~\ref{table:scaling-summary} where the DOS is scaled as $\exp(-(\epsilon_0/\epsilon)^{1/2})$, Efros concluded that the DC resistivity follows \Eq~(\ref{eq:rho-ES}), because the excitation spectrum of the particle screened by excitons that determines the DC transport still follows the power law $|E-E_\mathrm{F}|^{d-1}$~\cite{Efros76}.

{\bf\small (2) Energy region (D.1) in Table~\ref{table:scaling-summary}}

On the other hand, at high energies where the short-range part of the Coulomb interaction is dominant, the DOS follows the scaling of the soft Hubbard gap \Eq~(\ref{eq:DOS-HF}). Here we discuss the DC transport in the presence of the soft Hubbard gap without considering the long-range Coulomb interaction. First, we assume the DOS in the form
\begin{equation}
A(\epsilon) = \alpha |\epsilon|^\beta \exp(-(\gamma |\log\epsilon|)^d),\label{eq:sf-DOS}
\end{equation}
where $\epsilon=|E-E_\mathrm{F}|$. The power-law correction term $|\epsilon|^\beta$ is a slight generalization from \Eq~(\ref{eq:DOS-HF}). Under this assumption, we obtain the resistivity up to the leading order at low temperatures via variable-range hopping as
\begin{equation}
 \rho=\rho_0 \exp\left( c_0 \frac{\exp(-c_1| \log(k_\mathrm{B} T)|^{1/d})}{k_\mathrm{B} T}\right),\label{eq:rho-SH}
\end{equation}
for $d>1$, $c_0=1+a^{-1}(\frac{2\alpha}{\beta+1})^{-1/d}$, $a=\xi/2$ and $c_1>0$. For details of the derivation of \Eq~(\ref{eq:rho-SH}), readers are referred to Appendix~\ref{sec:appdx-transport1}.

{\bf\small (3) Energy region (D.4) in Table~\ref{table:scaling-summary}}

We next discuss the modification of the DC resistivity at energies lower than those justified by the ES scaling, by starting from \Eq~(\ref{eq:DOS-HF2}). We derive the temperature dependence of the DC resistivity in a way similar to the case of the soft Hubbard gap. Following a procedure similar to the case of the soft Hubbard gap (see Appendix~\ref{sec:appdx-transport2}), we obtain the resistivity up to the leading order as
\begin{eqnarray}
  \rho &=& \rho_0 \exp\left(\frac{(\beta/d)^{1/d} |\log(k_\mathrm{B} T)|^{-1/d}}{k_\mathrm{B} T}\right). \label{eq:rho-2}
\end{eqnarray}

\begin{figure}
  \begin{center}
  \includegraphics[width=0.375\textwidth,clip]{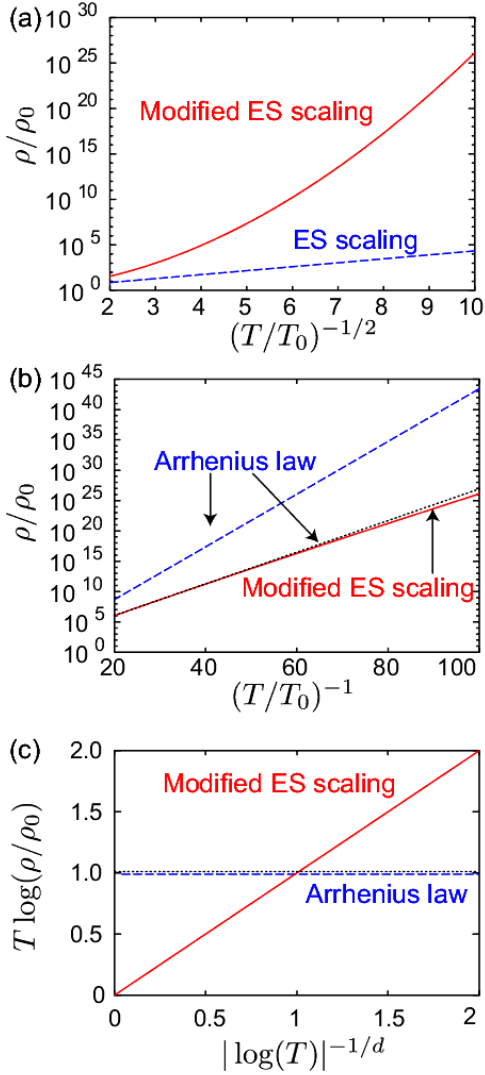}
  \end{center}
  \caption{\coloronline Comparison of the DC resistivity for the modified ES scaling, \Eq~(\ref{eq:rho-2}) with (a) that for the ES scaling, and (b) the Arrhenius law (the dashed and dotted lines). We note that the modified ES scaling shown by the solid line is hardly distinguishable from the Arrhenius law (the dotted line) by fitting the activation energy. In \Eq~(\ref{eq:rho-2}), we take $k_\mathrm{B}=1$ and $(\beta/d)^{1/d}=1$. (c) In this plot, all the data which follow the Arrhenius law converge to a constant, namely unity irrespective of the activation energy, while $\log(\rho/\rho_0)$ is proportional to $|\log(T)|^{-1/d}$ for the modified ES scaling. This distinguishes these two scaling laws clearly.}
  \label{fig:rho-comparison}
\end{figure}

The scaling laws of the DC resistivity obtained in the above discussion are summarized in Table~\ref{table:scaling-summary}. In Fig.~\ref{fig:rho-comparison} (a), we compare the obtained DC resistivity for the modified ES scaling, namely \Eq~(\ref{eq:rho-2}) with that for the ES scaling, namely \Eq~(\ref{eq:rho-ES}). Indeed, the DC resistivity for the modified ES scaling diverges at low temperatures faster than that for the ES scaling. However, as shown in Fig.~\ref{fig:rho-comparison} (b), the DC resistivity  for the modified ES scaling is almost indistinguishable from the Arrhenius law just by tuning the activation energy.

Here we propose a procedure to distinguish the modified ES scaling from the Arrhenius law clearly:
\begin{enumerate}
	\item One estimates $\rho_0$ in the scaling for the modified ES scaling (\Eq~(\ref{eq:rho-2})): From \Eq~(\ref{eq:rho-2}), one obtains
\begin{eqnarray}
 \log(\rho) &=& \log(\rho_0) + \frac{(\beta/d)^{1/d} |\log(k_\mathrm{B} T)|^{-1/d}}{k_\mathrm{B} T} \nonumber\\
            &\rightarrow& \log(\rho_0)~(1/T\rightarrow +0).
\end{eqnarray}
Thus one can estimate $\rho_0$ easily by a linear extrapolation of $\log(\rho)$ to the limit of $1/T\rightarrow +0$ with respect to $1/T$.

\item Next, one plots the logarithm of normalized resistivity multiplied by the temperature $T\log(\rho/\rho_0)$ against $|\log(T)|^{-1/d}$ as shown in Fig.~\ref{fig:rho-comparison} (c). In this plot, $\log(\rho/\rho_0)$ is proportional to $|\log(T)|^{-1/d}$ for the modified ES scaling, while the normalized resistivity $\log(\rho/\rho_0)$ is constant when the experimental data follows the Arrhenius law. These two functions can be distinguished clearly from each other.
\end{enumerate}
We believe that the validity of the present theory can be tested against experiments by using this plot provided that $\rho_0$ does not contain other extrinsic factors of temperature dependence out of the present consideration, such as the volume expansion.

\subsection{Comparison with experiments}
\begin{figure}
  \begin{center}
  \includegraphics[width=0.39\textwidth,clip]{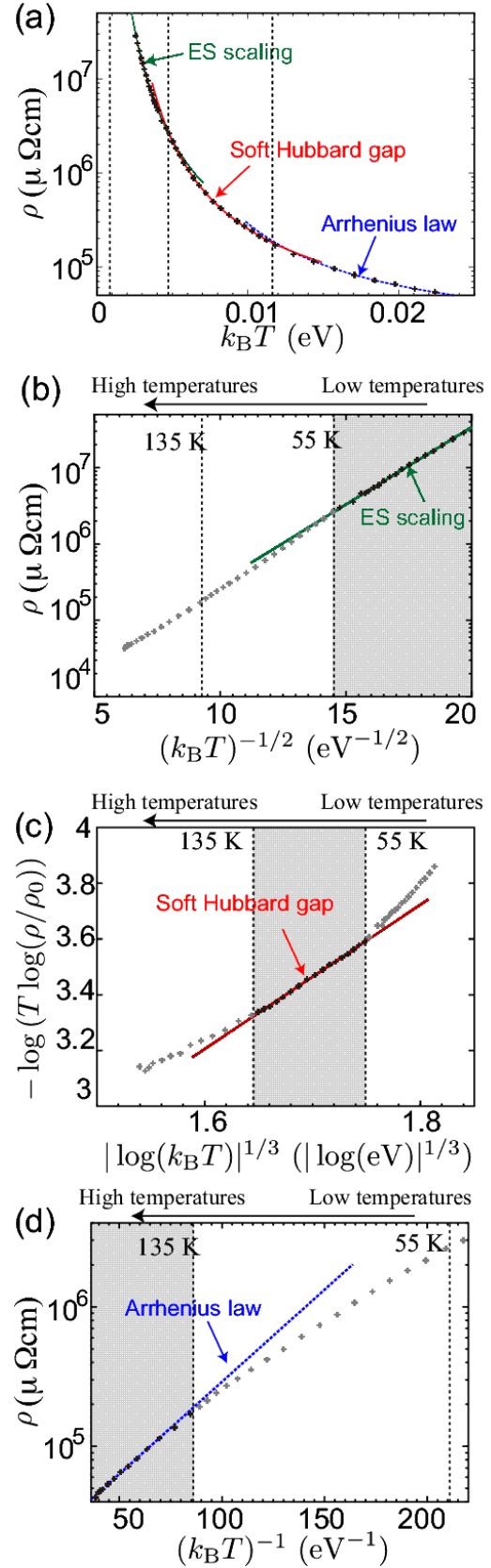}
  \end{center}
	\caption{\coloronline Scaling analyses of the DC resistivity of SrRu$_{1-x}$Ti$_x$O$_3$ (the experimental data illustrated as filled circles are from Ref.~\citen{Kim05}). (a) The logarithmic plot. The panels (b), (c) and (d) show scaling plots for the ES scaling ($T<55 \mbox{K}$), the soft Hubbard gap ($55 \mbox{K}<T<135 \mbox{K}$) and the Arrhenius law ($135 \mbox{K}<T$), respectively. The vertical dotted lines denote $T=55 ~\mbox{K}, 135~\mbox{K}$.}
  \label{fig:STRO}
\end{figure}
In this section, we compare our formulae for the DC resistivity, \Eq~(\ref{eq:rho-SH}) and \Eq~(\ref{eq:rho-2}) with transport properties of SrRu$_{1-x}$Ti$_x$O$_3$~\cite{Kim05}, where the breakdown of the ES scaling was indicated in photoemission experiments~\cite{Kim06, Maiti07}. In Fig.~\ref{fig:STRO}, we show the experimental data of the DC resistivity of SrRu$_{1-x}$Ti$_x$O$_3$ at $x = 0.6$, where the ground state is insulating. The experimental data fit well with the ES scaling for $k_\mathrm{B} T < 0.005~\mathrm{eV}~(T \lesssim 60~\mathrm{K})$, indicating that the ground state is an Anderson insulator. At higher temperatures ($0.005~\mathrm{eV} < k_\mathrm{B}T< 0.011~\mathrm{eV},~60~\mathrm{K}\lesssim T\lesssim 130~\mathrm{K}$), however, they fit well with the scaling obtained in the presence of the soft Hubbard gap for 3D (\Eq~(\ref{eq:rho})), being consistent with our theory. The fitting parameters are as follows: $\log(\rho_0) = 9.0\pm 0.2$, $\log(c_0) = 0.94\pm 0.7$ and  $c_1 = 2.6 \pm 0.4$. On the other hand, for higher temperatures ($0.011~\mathrm{eV} < k_\mathrm{B}T$), the experimental data deviate from the scaling for the soft Hubbard gap, being consistent with the our 3D HF study where the asymptotic behavior of the soft Hubbard gap is restricted to low energies ($<0.1 t$) as shown in Figs.~\ref{fig:DOS} and \ref{fig:dos-scaling-HF-3D}. On the other hand, we cannot find the region where the modified ES scaling predicted by \Eq~(\ref{eq:rho-2}) is satisfied even in the lowest temperature available in the experiment ($\simeq 30~\mathrm{K}$). We believe that further experiments at lower temperature will reveal the existence of the correction to the ES scaling.

Another interpretation of the experimental data for $0.005~\mathrm{eV} < k_\mathrm{B}T< 0.011~\mathrm{eV}$ may be that the ES scaling ($k_\mathrm{B}T < 0.005~\mathrm{eV}$) and the Arrhenius law ($k_\mathrm{B}T>0.011~\mathrm{eV}$) cross over there; the Arrhenius law arises from the hopping conduction between nearest-neighbor localized states, which dominates the DC resistivity at high temperatures instead of the variable-range hopping. Indeed, for $0.011~\mathrm{eV}<k_\mathrm{B}T< 0.026~\mathrm{eV}~(130~\mathrm{K} \lesssim T \lesssim 300~\mathrm{K})$, they seem to follow the Arrhenius law with a small activation energy $0.030\pm0.001~\mathrm{eV}$. In order to exclude or support the possibility of this crossover, further experiments as functions of composition $x$ in the vicinity of the metal-insulator transition are desired.

%% file: conclusion.tex
\section{Summary and discussion}\label{SEC:CONCLUSION}
\begin{table*}[t]
 \centering
 \begin{tabular}{l!{\bvline{1.5pt}}lll}\bhline{1.5pt}
  \multicolumn{1}{c!{\bvline{1.5pt}}}{{\small Model}} & {\small Scaling of DOS} & {\small DC resistivity} & {\small \!\!\!\!Temperature}\\ \bhline{1.5pt}
  {\small Clean Mott insulator} & $=0$ (hard gap)&  $\exp(T_0/T)$ &\\ \bhline{1.5pt}
  {\small (A) Short-range interaction } &$\simeq A_0>0$ & $\exp((T_0/T)^{1/(d+1)})$&\\ 
  {\small \textit{without} multivalley energy landscape}  & & &\\\hline
  {\small (B) Short-range interaction} &\cellcolor{lightgray} $\exp\left({-\gamma(-\log \epsilon)^d}\right)$ & \cellcolor{lightgray}$\exp\left( c_0 \frac{\exp(-c_1| \log(k_\mathrm{B} T)|^{1/d})}{k_\mathrm{B} T}\right)$ &\\
	{\small \textit{with} multivalley energy landscape}  &~~~~$\cdots$ \Eq~(\ref{eq:DOS-HF}) &~~~~$\cdots$ \Eq~(\ref{eq:rho-SH})&\\ \bhline{1.5pt}
  {\small (C) Long-range Coulomb interaction}&(C.1) $\epsilon^{d-1}$ & $\exp\left((T_0/T)^{1/2}\right)$ &(HTs)\\
  {\small \textit{without} multivalley energy landscape}    &(C.2) $\exp(-(\epsilon_0/\epsilon)^{1/2})~(\mbox{3D})$ & $\exp\left((T_0/T)^{1/2}\right)$ &(LTs)\\ \hline
                          &\cellcolor{lightgray}(D.1) $\exp\left({-\gamma(-\log \epsilon)^d}\right)$ & \cellcolor{lightgray}$\exp\left( c_0 \frac{\exp(-c_1| \log(k_\mathrm{B} T)|^{1/d})}{k_\mathrm{B} T}\right)$ &(HTs) \\ 
													&~~~~$\cdots$ \Eq~(\ref{eq:DOS-HF}) &~~~~$\cdots$ \Eq~(\ref{eq:rho-SH})& $\ \ \ \downarrow$\\  \cdashline{2-3}[.4pt/1pt]
  {\small (D) Long-range Coulomb interaction}&(D.2) $\epsilon^{d-1}$ & $\exp\left((T_0/T)^{1/2}\right)$ & $\ \ \ \downarrow$\\
  {\small \textit{with} multivalley energy landscape}    &(D.3) $\exp(-(\epsilon_0/\epsilon)^{1/2})~(\mbox{3D})$ & $\exp\left((T_0/T)^{1/2}\right)$ & $\ \ \ \downarrow$\\
  &\cellcolor{lightgray}(D.4) $\exp(-\beta \epsilon^{-d})$ & \cellcolor{lightgray}$\exp\left(c_0 \frac{|\log(k_\mathrm{B} T)|^{-1/d}}{k_\mathrm{B} T}\right)$ &(LTs) \\
	&~~~~$\cdots$ \Eq~(\ref{eq:DOS-HF2}) &~~~~$\cdots$ \Eq~(\ref{eq:rho-2})&\\ \bhline{1.5pt}
 \end{tabular}
 \caption{Summary of scaling laws of DOS and DC resistivity for four kinds of models: (A)/(B) short-range interaction without/with a multivalley energy landscape, (C)/(D) long-range interaction without/with a multivalley energy landscape. Shaded part denotes the novel scaling laws obtained in this paper. The scaling of the case (C.2) is expected to appear in energies lower than that of the case (C.1). The case (D.1) corresponds to the energy scale where the short-range part of the Coulomb interaction is dominant, while the cases (D.2-4) correspond to that where the long-range part is dominant. In the cases of the long-range interaction, scaling crossovers are expected as a function of the temperature. Abbreviations are: High temperatures (HTs); Low temperatures (LTs).}
 \label{table:scaling-summary}
\end{table*}
In summary, we have examined the ground state and single-particle excitations of the 3D Anderson-Hubbard model within the Hartree-Fock approximation. In \S~3, although only the short-range interaction is present in the model in contrast to the Efros-Shklovskii theory, we have found an unconventional soft gap over the whole insulating phases. Namely, we have found that the DOS vanishes toward the Fermi energy faster than any power law indicating the formation of an unconventional soft gaps, which we call \textit{soft Hubbard gap}. Further numerical evidences to support this unconventional soft gap has been given within the Hartree-Fock approximation, and further with the exact diagonalization in 1D. In contrast to the 3D case, a power-law scaling is satisfied in 1D.

In \S~4, based on the picture of a multivalley energy landscape, we have constructed a phenomenological theory, being qualitatively as well as quantitatively consistent with the numerically-obtained scaling of the soft Hubbard gap. There, further support for the scaling theory has been given in 2D within the Hartree-Fock approximation. Moreover, by considering effects of the long-range part of the Coulomb interaction remaining unscreened at low energies, we have obtained a novel scaling of the soft gap beyond the Efros-Shklovskii theory. This novel scaling is caused by a multivalley energy landscape. Finally, we have proposed scaling crossovers of the DOS as the long-range part becomes dominant at low temperatures. Scaling laws and their expected crossovers are summarized in Table~\ref{table:scaling-summary}.

In \S~5, we have derived the temperature dependence of the DC resistivity in the presence of the soft gaps. Possible experiments to verify the present theory has been proposed. The scaling laws of the density of states and the DC resistivity are summarized in Table~\ref{table:scaling-summary}. Finally, we have shown that the present theory is indeed consistent with the experimental data for SrRu$_{1-x}$Ti$_x$O$_3$.

In order to readdress the essence of the present theory and stimulate further theoretical studies, we now discuss possible extension of the present theory. In \S~4, we have shown that a gapless collective mode accompanying a multivalley energy landscape causes relaxations from a single-particle excited states to a multiply-excited states not counted in the single-particle DOS, which directly leads to the formation of the unconventional soft gap. In other words, the single-particle excited states are eliminated from low energies by the ground-state stability against \textit{the collective excitations} at low energies. However, there are actually many other kinds of gapless collective modes. A typical example is a spin wave accompanying magnetic ordering with spontaneous symmetry breaking, which is not included in the present Hartree-Fock approximation. If these collective modes also cause the relaxation from a single-particle excited states to a multiply-excited states, the present scaling of the density of states will equally be valid, which extends the applicable scope of the present theory. Whether or not the present theory can be extended to general collective modes is left for a future challenge.

Recently \textit{magnetic hard gaps} have been observed in a large number of disordered materials such as In-doped CdMnTe~\cite{Terry92a, Terry92b}, amorphous GeCr-films~\cite{Aleshin88}, doped Si~\cite{Sarachik85} and amorphous Si$_{1-x}$Mn$_x$-films~\cite{Yakimov95} by measuring the temperature dependence of the DC resistivity. These gaps were identified to be of magnetic origin by reduction of the gap sizes by an increased external magnetic field. Because the DOS vanishes toward the Fermi energy even faster than a power law in the presence of the soft Hubbard gap, the soft Hubbard gap causes the rapid divergence of the DC resistivity slightly slower than the Arrhenius law at low temperatures as summarized in Table~\ref{table:scaling-summary}. Comparisons of the present theory with these experiments are interesting issues.

%% file: ackn01.tex
\section*{Acknowledgments}
We thank B. Shklovskii and T. Ohtsuki for fruitful discussions. M. I. thanks Aspen Center for Physics for the hospitality. Numerical calculation was partly carried out at the Supercomputer Center, Institute for Solid State Physics, Univ. of Tokyo. This work is financially supported by MEXT under the grant numbers 16076212, 17071003 and 17064004. H. S. thanks JSPS for the financial support.

%% file: ap01.tex
\section{Derivation of temperature dependence of DC resistivity}\label{sec:appdx-transport}
In this appendix, we derive the temperature dependence of the DC resistivity in the presence of the soft Hubbard gap or the soft Coulomb gap within the variable-range hopping.
\subsection{Soft Hubbard gap: Energy region (D.1) in Table~\ref{table:scaling-summary}}\label{sec:appdx-transport1}
First, we derive the DC transport in the presence of the soft Hubbard gap. We assume the DOS in the form
\begin{equation}
	A(E) = \alpha |\epsilon|^\beta \exp(-(\gamma |\log\epsilon|)^d),\label{eq:sf-DOS}
\end{equation}
where $\epsilon=|E-E_\mathrm{F}|$. The power-law correction term $|\epsilon|^\beta$ is a slight generalization from \Eq~(\ref{eq:DOS-HF}). In the following, we discuss $\rho$ for the soft Hubbard gap without considering the long-range Coulomb interaction.

In order to determine the energy window $\epsilon_0$ dominating the transport for a given temperature $T$, we minimize the resistivity
\begin{equation}
 \rho = \rho_0 \exp\left( \frac{1}{N^{1/d}(\epsilon_0) a} + \frac{\epsilon_0}{k_\mathrm{B} T} \right), \label{eq:rho}
\end{equation}
derived from the hopping transport within the energy window of the width $\epsilon_0$, where $a=\xi/2$. We define the number of electrons within the energy window, $N(\epsilon_0)$ as
\begin{eqnarray}
	N(\epsilon_0) &=& \int_{-\epsilon_0+E_\mathrm{F}}^{\epsilon_0+E_\mathrm{F}} A(E) d E \nonumber\\
 &=& 2\alpha \int_{c}^{+\infty} \exp(-\gamma^d x^d -(\beta+1) x) d x \nonumber \\
 &\simeq& \frac{2\alpha}{\beta+1} \exp(-\gamma^d c^d -(\beta+1) c)  \nonumber \\
 &=& \frac{2\alpha}{\beta+1} \epsilon_0^{\beta+1} \exp(-\gamma^d |\log \epsilon_0|^d),\label{eq:N}
\end{eqnarray}
where $x=-\log(\epsilon)$, $c=-\log(\epsilon_0)$. Here we employ an approximation
\begin{eqnarray}
  \int_{c}^{+\infty} \exp(-ax^d-x) dx\!\!&=&\!\!\int_c^\infty \sum_{n=0}^{\infty} \frac{1}{n!} (-a)^n  (x^d)^n e^{-x} d x\nonumber \\
  &=& \sum_{n=0}^{\infty} \frac{1}{n!} (-a)^n \Gamma(1+dn, c) \nonumber \\
  &\simeq& \sum_{n=0}^{\infty} \frac{1}{n!} (-a)^n c^{dn} e^{-c} \nonumber \\
  &=& \exp(-ac^d-c),\label{eq:int}
\end{eqnarray}
which is justified for $c\gg1$. In the fourth line, we expand the incomplete gamma function $\Gamma (z,p)$ in $p$ as
\begin{eqnarray}
\!\!\!\!\Gamma(z,p)&=&p^{z-1} e^{-p} \nonumber \\
\!\!&\times&\!\!\!\!\!\left\{\!1\!+\!\sum_{n=1}^\infty \frac{1}{p^n} (z-1)(z-2) \cdots (z-n) \right\}.
\end{eqnarray}

By differentiating \Eq~(\ref{eq:rho}) with respect to $\epsilon_0$, we obtain
\begin{eqnarray}
 \frac{d \rho}{d \epsilon_0} &=& -\left\{ \frac{2A(\epsilon_0) }{ad N^{1/d+1}(\epsilon_0)} - \frac{1}{k_\mathrm{B} T}\right\} \rho(\epsilon_0).
\end{eqnarray}
Thus the minimization condition of the resistivity is given by
\begin{equation}
 \frac{2A(\epsilon_0) }{ad N^{1/d+1}(\epsilon_0)} =  \frac{1}{k_\mathrm{B} T}.\label{eq:minimum-condition}
\end{equation}
By taking logarithms of both sides of \Eq~(\ref{eq:minimum-condition}) and substituting \Eqs~(\ref{eq:sf-DOS}) and (\ref{eq:N}), we obtain
\begin{equation}
 \gamma^d |\log \epsilon_0|^d + \left(\beta+d+1\right)  |\log \epsilon_0|=d|\log(k_\mathrm{B} T)|,\label{eq:determine_epsilon_0}
\end{equation}
up to the next leading term. In general, a solution of an equation for $x$, 
\begin{equation}
  a_0 x+ b_0x^{1/d} = c_0
\end{equation}
is given by 
\begin{equation}
  x=c_0 a_0^{-1} - b_0 c_0^{1/d} a_0^{-1-1/d} + O(c_0^{-1+2/d}). 
\end{equation}
Thus, for $d>1$, we obtain the solution of \Eq~(\ref{eq:determine_epsilon_0}) up to the next leading term in the low-temperature limit as
\begin{eqnarray}
 \gamma^d |\log \epsilon_0|^d &=& d X - (\beta+d+1)d^{1/d}\gamma^{-1}X^{1/d}\nonumber\\
 &&+O(X^{-1+2/d}), 
\end{eqnarray}
\begin{eqnarray}
 \epsilon_0 &=& \exp\Bigl(-d^{1/d} \gamma^{-1} X^{1/d}\nonumber\\
 &&+(\beta+d+1) d^{-2+2/d} \gamma^{-2}X^{-1+2/d}\nonumber\\
 &&+O\bigl(X^{-2+3/d}\bigr)\Bigr),
\end{eqnarray}
where $X$ represents $|\log(k_\mathrm{B} T)|$. By substituting these equations into \Eq~(\ref{eq:N}), we obtain
\begin{equation}
 N(\epsilon_0) =  \frac{2\alpha}{\beta+1}  (k_\mathrm{B} T)^d \exp(d^{1/d+1}\gamma^{-1}X^{1/d}).
\end{equation}
Thus the first term in the exponential function in \Eq~(\ref{eq:rho}) up to the leading term is given by
 \begin{eqnarray}
  \frac{1}{N(\epsilon_0)^{1/d} a} &=& \frac{\exp(-d^{1/d}\gamma^{-1}X^{1/d})}{a\left(\frac{2\alpha}{\beta+1}\right)^{1/d} k_\mathrm{B} T}.
 \end{eqnarray}
 On the other hand, the second term in the exponential function up to the leading term in \Eq~(\ref{eq:rho}) is given by
  \begin{eqnarray}
  \frac{\epsilon_0}{k_\mathrm{B} T} &=& \frac{\exp(-d^{1/d}\gamma^{-1}X^{1/d})}{k_\mathrm{B} T}.
 \end{eqnarray}
 Thus we obtain the resistivity up to the leading order at low temperatures as
\begin{equation}
 \rho=\rho_0 \exp\left( c_0 \frac{\exp(-c_1| \log(k_\mathrm{B} T)|^{1/d})}{k_\mathrm{B} T}\right),
\end{equation}
for $d>1$, $c_0=1+a^{-1}(\frac{2\alpha}{\beta+1})^{-1/d}$ and $c_1>0$. Equation~(\ref{eq:rho-SH}) gives the scaling of the resistivity for our soft Hubbard gap (given by \Eq~(\ref{eq:DOS-HF})) obtained for the model with the short-range interaction only.

\subsection{Modified ES scaling: Energy region (D.4) in Table~\ref{table:scaling-summary}}\label{sec:appdx-transport2}
We next discuss the modification of the DC resistivity at energies lower than those justified by the ES scaling, by starting from \Eq~(\ref{eq:DOS-HF2}). We derive the temperature dependence of the DC resistivity in a way similar to the case of the soft Hubbard gap.

We integrate the DOS with respect to energy using the expansion of the incomplete gamma function and obtain
\begin{eqnarray}
	N(\epsilon_0) &=& \int_{-\epsilon+E_\mathrm{F}}^{\epsilon_0+E_\mathrm{F}} A(E) d E \nonumber \\
 &=& 2\alpha \beta^{-1} d^{-1} \Gamma(-\frac{1}{d}, \beta \epsilon_0^{-d}) \nonumber\\
 &\simeq& 2\alpha \beta^{-1} d^{-1} \epsilon_0^{d+1} \exp(-\beta |\epsilon_0|^{-d}). \label{eq:N-2}
\end{eqnarray}

By taking logarithms of both sides of \Eq~(\ref{eq:minimum-condition}), we obtain
\begin{equation}
 \beta d^{-1} \epsilon_0^{-d} +(1+1/d)^2 \log (\epsilon_0^{-d}) = |\log(k_\mathrm{B} T)|. \label{eq:determine_epsilon_0-2}
\end{equation}
up to the next leading term. In general, the solution of an equation for $x$ in the limit of $c_0\rightarrow +\infty$, 
\begin{eqnarray}
    a_0 x+ b_0 \log(x) &=& c_0
\end{eqnarray}
is given by 
\begin{eqnarray}
   x &=& c_0 a_0^{-1} - a_0^{-1}b_0 \log(c_0) + O(c_0^0).
\end{eqnarray}
Thus we obtain the solution of \Eq~(\ref{eq:determine_epsilon_0-2}) up to the next leading term in the low-temperature limit as
\begin{eqnarray}
 \epsilon_0^{-d} &=& d\beta^{-1}X-d\beta^{-1}(1+1/d)^2 \log (X) \nonumber \\
 && + O(X^0), \\
 \epsilon_0 &=&  (\beta/d)^{1/d}X^{-1/d}  \nonumber\\
  && +(\beta/d)^{1/d} d^{-1} (1+1/d)^2 X^{-(1+1/d)} \log(X)\nonumber\\
  && + O(X^{-(1+1/d)}),
 \end{eqnarray}
where $X$ represents $|\log(k_\mathrm{B} T)|$ again. Thus the first term in the exponential function in \Eq~(\ref{eq:rho}) up to the leading term is given by
 \begin{eqnarray}
  \frac{1}{N(\epsilon_0)^{1/d} a} &=& (C/a) \frac{|\log(k_\mathrm{B} T)|^{-(1+1/d)}}{k_\mathrm{B} T},
 \end{eqnarray}
 where $C$ is a constant. On the other hand, the second term in the exponential function up to the leading term in \Eq~(\ref{eq:rho}) is given by
  \begin{eqnarray}
  \frac{\epsilon_0}{k_\mathrm{B} T} &=& \frac{(\beta/d)^{1/d} |\log(k_\mathrm{B} T)|^{-1/d}}{k_\mathrm{B} T}.
 \end{eqnarray}
 Thus we obtain the resistivity up to the leading order as
\begin{eqnarray}
  \rho &=& \rho_0 \exp\left(\frac{(\beta/d)^{1/d} |\log(k_\mathrm{B} T)|^{-1/d}}{k_\mathrm{B} T}\right).
\end{eqnarray}